\documentclass[aps,prd,twocolumn,preprintnumbers,superscriptaddress,floatfix,nofootinbib,notitlepage,showkeys,showpacs]{revtex4-1}

\usepackage[utf8x]{inputenc}
\usepackage{titlesec}
\usepackage{graphicx}
\usepackage{hyperref}
\usepackage{latexsym}
\usepackage{amsmath}
\usepackage{amssymb}
\usepackage{bbm}
\usepackage{ulem}
\usepackage{pdfsync}
\usepackage{epsfig}
\usepackage{epstopdf}
\usepackage{color}
\usepackage{comment}
\usepackage{placeins}
\usepackage{slashed}
\usepackage{mathrsfs}

\newcommand{\ie}{{\it i.e.}}
\newcommand{\eg}{{\it e.g.\,}}
\newcommand{\Msun}{M_{\odot}}

\newcommand{\gsim}{\lower.7ex\hbox{$\;\stackrel{\textstyle>}{\sim}\;$}}
\newcommand{\lsim}{\lower.7ex\hbox{$\;\stackrel{\textstyle<}{\sim}\;$}}

\begin{document}
%\rightline{CERN-TH-2018-053}

\title{Constraining Primordial Black Holes with the EDGES 21-cm Absorption Signal}

\author{Andi Hektor}
\email{andi.hektor@cern.ch}
\affiliation{NICPB, R\"avala 10, 10143 Tallinn, Estonia}

\author{Gert H\"utsi}
\email{gert.hutsi@to.ee}
\affiliation{Tartu Observatory, University of Tartu, Observatooriumi 1, 61602 T\~oravere, Estonia}
\affiliation{NICPB, R\"avala 10, 10143 Tallinn, Estonia}

\author{Luca Marzola}
\email{luca.marzola@cern.ch}
\affiliation{NICPB, R\"avala 10, 10143 Tallinn, Estonia}

\author{Martti Raidal}
\email{martti.raidal@cern.ch}
\affiliation{NICPB, R\"avala 10, 10143 Tallinn, Estonia}
\affiliation{Theoretical Physics Department, CERN, Geneva, Switzerland}

\author{Ville Vaskonen}
\email{ ville.vaskonen@kbfi.ee}
\affiliation{NICPB, R\"avala 10, 10143 Tallinn, Estonia}

\author{Hardi Veerm\"ae}
\email{hardi.veermae@cern.ch}
\affiliation{Theoretical Physics Department, CERN, Geneva, Switzerland}
\affiliation{NICPB, R\"avala 10, 10143 Tallinn, Estonia}

\begin{abstract}
\noindent
The EDGES experiment has recently measured an anomalous global 21-cm spectrum due to hydrogen absorptions at redshifts of about $z\sim 17$. Model independently, the unusually low temperature of baryons probed by this observable sets strong constraints on any physical process that transfers energy into the baryonic environment at such redshifts. Here we make use of the 21-cm spectrum to derive bounds on the energy injection due to a possible population of ${\cal O}(1-100) M_\odot$ primordial black holes, which induce a wide spectrum of radiation during the accretion of the surrounding gas. After calculating the total radiative intensity of a primordial black hole population, we estimate the amount of heat and ionisations produced in the baryonic gas and compute the resulting thermal history of the Universe with a modified version of RECFAST code. Finally, by imposing that the temperature of the gas at $z\sim 17$ does not exceed the indications of EDGES, we constrain the possible abundance of primordial black holes. Depending on uncertainties related to the accretion model, we find that ${\cal O}(10) M_\odot$ primordial black holes can only contribute to a fraction $f_{\rm PBH}<(1-10^{-3})$ of the total dark matter abundance.
\end{abstract}

\maketitle

%===============================================================================
% BODY
%===============================================================================

%-------------------------------------------------------------------------------
\section{Introduction }
\label{sec:introduction}
%-------------------------------------------------------------------------------

The Experiment to Detect the Global Epoch of Reionization Signature (EDGES) has recently found an anomalously strong absorption in the 21-cm spectrum by the baryonic gas at redshifts in the range $z\approx 15-21$~\cite{Bowman:2018aa}.  Because the intensity of the detected signal is proportional to $I_{\rm 21cm}\propto 1 - (T_R(z)/T_S(z)),$ where $T_S$ is the spin temperature of the atomic hydrogen and $T_R$ is the temperature of background radiation, the very first studies of the EDGES anomaly ascribed the extra absorption to a new cooling mechanism for the hydrogen gas, based on baryon-dark matter (DM) interactions~\cite{Barkana:2018aa, Fialkov:2018xre, Munoz:2018pzp, Berlin:2018sjs,Barkana:2018qrx,Kang:2018qhi} (this idea was studied before the EDGES anomaly appeared in Refs.~\cite{Dvorkin:2013cea,Tashiro:2014tsa,Munoz:2015bca}), on modified onset of star formation~\cite{Mirocha:2018cih,Safarzadeh:2018hhg}, or on the effects of dark energy~\cite{Hill:2018lfx,Costa:2018aoy}. DM annihilations~\cite{Fraser:2018acy,DAmico:2018sxd,Yang:2018gjd} and black hole accretion~\cite{Ewall-Wice:2018bzf}, however, result in the injection of particles characterised by a broad energy spectrum that inevitably lead to heating the hydrogen clouds. An alternative solution that does not present this problem considers the injection of soft photons to increase $T_R$ consistently with the Cosmic Microwave Background (CMB) constraints~\cite{Fraser:2018acy,Pospelov:2018kdh}. Interestingly, the presence of an extra photon background could also be supported by the ARCADE~2~\cite{2011ApJ...734....5F} excess, as predicted in~\cite{Feng:2018rje}.

In spite of the mechanism responsible for its generation, the EDGES signal can be used to constrain all new physics scenarios that result in energy injection into the baryonic environment, thereby heating the hydrogen clouds responsible for the absorption signal. This observation was used for example in~\cite{DAmico:2018sxd} to constrain DM annihilations. In this work we adopt the same attitude to analyze the energy injection due to a population of ${\cal O}(1-100) M_\odot$ primordial black holes (PBHs), with the aim of constraining the possible PBH DM abundance. To this purpose we disregard other sources of radiation, neglecting for instance all the astrophysical processes ongoing in the early Universe. The results obtained in this work should therefore be regarded as conservative.

PHBs are among the oldest candidates to explain the observed DM abundance~\cite{Hawking:1971ei,Carr:1974nx,Carr:1975qj,1975A&A....38....5M,CHAPLINE:1975aa}. The detection of gravitational waves from black holes mergers with mass ${\mathcal O}(10)\Msun$ by the LIGO and VIRGO interferometers~\cite{Abbott:2016blz,Abbott:2016nmj} has recently revived the interest of the community in this topic~\cite{Carr:2016drx,Kashlinsky:2016sdv,Bird:2016dcv,Clesse:2016vqa,Sasaki:2016jop} (for a review see, \eg, \cite{Sasaki:2018dmp}). At the present stage, the accumulated experimental data collectively constrain the fraction of PBH DM, $f_{\rm PBH}$, to be below unity~\cite{Carr:2017jsz,Sasaki:2018dmp}, barring scenarios in which the radiation induced by PBH is strongly modified~\cite{Raidal:2018eoo} or misinterpretations of the results of lensing experiments~\cite{Garcia-Bellido:2017xvr,Clesse:2017bsw,Garcia-Bellido:2017imq}.

To further investigate the matter, we estimate the effects of a PBH population on the kinetic temperature of the cosmic gaseous medium, resulting from the wide spectrum of radiation produced by the cosmic gas accreting onto PBHs.  We  calculate the total radiative intensity of the PBH population and estimate the fraction of this energy absorbed by the gas, which induces additional source terms in the set of differential equations that describe the ionization and the temperature evolution of the cosmic gas. By limiting the amount gas heating at $z\sim 17$ in accordance to the EDGES measurement, we then compute the maximal fraction of PBH DM allowed. Our approach is therefore similar to the ones previously adopted in the literature to constrain the PBH abundance from CMB observations~\cite{Ricotti:2007au,Horowitz:2016lib,Ali-Haimoud:2016mbv,Poulin:2017bwe}.

The paper is organized as follows. In Sec.~\ref{sec:formalism_assumption} we provide a brief description of the adopted formalism, introducing the gas accretion model and the spectral templates adopted in the calculation of the the PBHs total radiative intensity. After that we describe the impact of the latter on the evolution equations for the ionization and thermal history of the cosmic baryonic medium. In Sec.~\ref{sec:results} we present our main results, which we summarise and discuss in Sec.~\ref{sec:conclusions}. In the appendix we calculate approximate CMB bounds for the considered parametrized accretion model, and compare them to the results obtained from the 21-cm spectrum.

%-------------------------------------------------------------------------------
\section{Adopted formalism and accretion model}
\label{sec:formalism_assumption}
\subsection{Energy injection from PBHs}
%-------------------------------------------------------------------------------

\begin{figure}[t]
\begin{center}
\includegraphics[width=.9\linewidth]{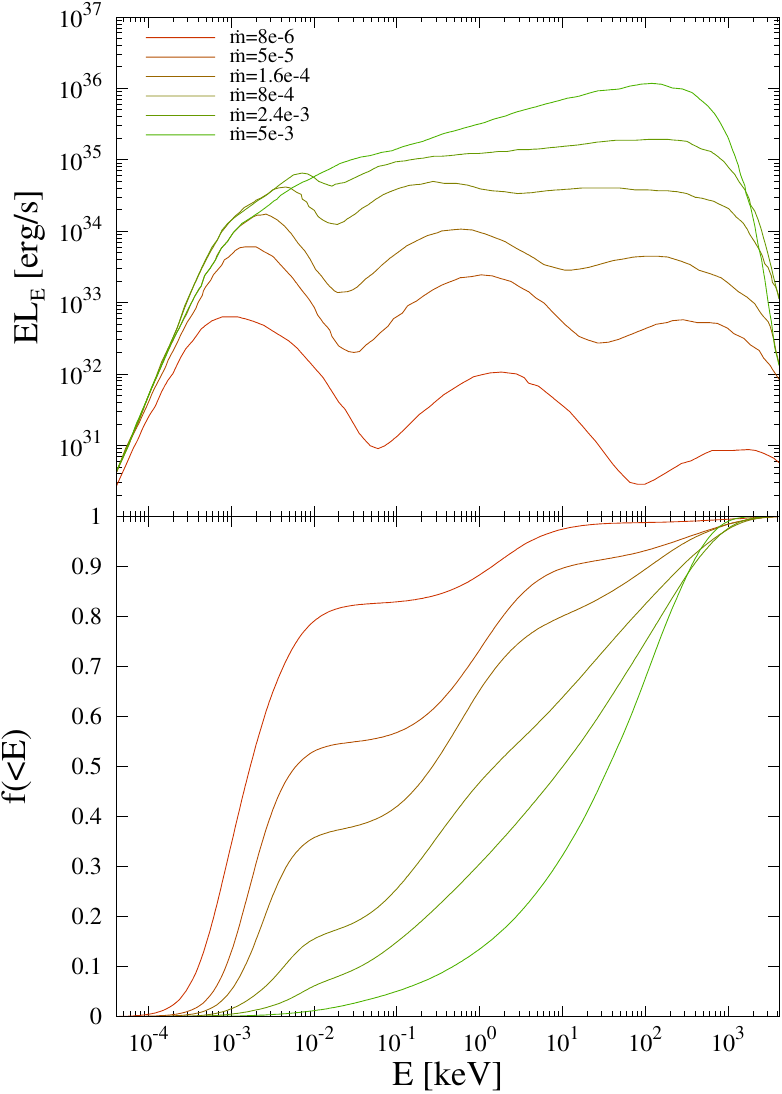}
\caption{Upper panel: The ADAF model spectra for different values of the specific accretion rate $\dot{m}\equiv \dot{M}/\dot{M}_{\rm Edd}$~\cite{2014ARA&A..52..529Y}. Lower panel: Cumulative energy distributions for the above ADAF models.}
\label{fig:ADAF_template}
\end{center}
\end{figure}

\begin{figure}[t]
\centering
\includegraphics[width=.96\linewidth]{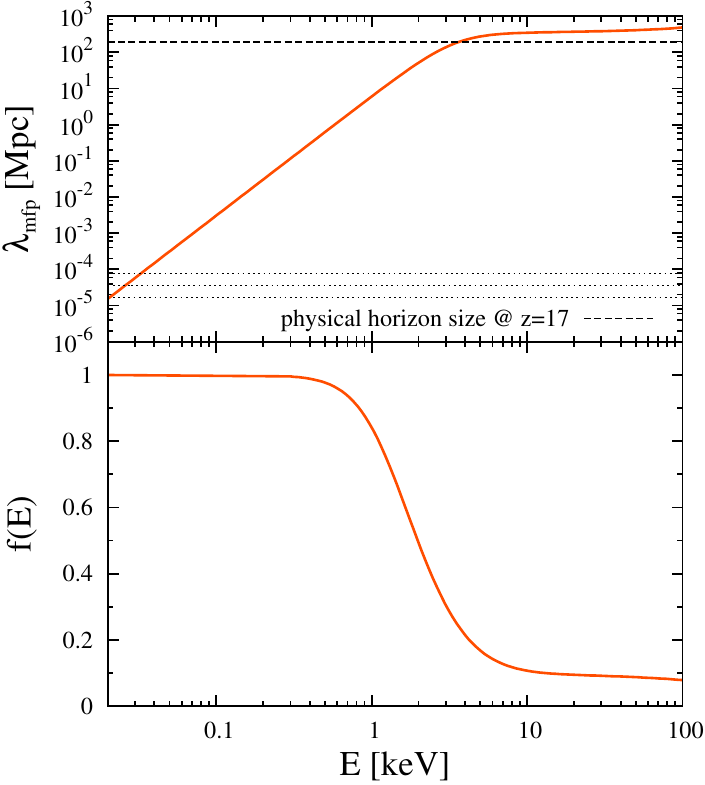}
\caption{Upper panel: Mean free path in physical units for photons observed at redshift $z=17$. The dashed horizontal line indicates the horizon size at this epoch. Thin dotted lines show the mean separation between PBHs with masses $1$, $10$ and $100\Msun$, assuming $f_{\rm PBH}=1$. Lower panel: The corresponding `on-spot' correction factor.}
\label{fig:mfp_f_param}
\end{figure}

The gas surrounding black holes produces a wide spectrum of radiation during the gravitational infall that drives the accretion process. The Bondi accretion model~\cite{Bondi:1952ni,2002apa..book.....F} provides a useful starting point to estimate the influx of the gas, but the Bondi mass accretion rate $\dot M_B$ must first be corrected by a factor $\lambda$ to be consistent with the non-observation of a significant population of isolated Galactic neutron stars. The present upper bound on $\lambda$ is $\sim 10^{-2}-10^{-3}$~\cite{Perna:2003ck}.
 
To estimate the emitted photon flux, we further modify the Bondi model to account for the radiative efficiency. As we are interested in a regime characterised by low mass accretion rate and opacity, we adopt the advection dominated accretion flow (ADAF) model~\cite{Narayan:1994xi}.
Under these assumptions, the radiative efficiency $\eta$, corresponding to a bolometric luminosity $L=\eta \dot{M}c^2$, can be approximated as
\begin{equation}\label{eq:eta}
\eta = 0.1 \times
\begin{cases}
\left(\frac{\dot{m}}{\dot{m}_{\rm crit}}\right)^\beta & \text{if $\dot{m} \le \dot{m}_{\rm crit}$} \\
1 & \text{if $\dot{m} > \dot{m}_{\rm crit}$}\,,
\end{cases}
\end{equation}
where $\dot{m}_{\rm crit}\simeq 0.01$, $\beta \sim 0-1$~\cite{Narayan:2008bv,2014ARA&A..52..529Y} and $\dot{m}$ is the mass accretion rate in units of the Eddington rate: $\dot{m}\equiv \dot{M}/\dot{M}_{\rm Edd}$. The resulting specific luminosity is then~\footnote{It turns out that the condition $\dot{m} \le \dot{m}_{\rm crit}$ is always satisfied.}
\begin{equation}\label{eq:lumi}
L_E(z) = L_{\rm Edd}\,\dot{m}(z) \left(\frac{\dot{m}(z)}{\dot{m}_{\rm crit}}\right)^\beta \times f(E),
\end{equation}
where $L_{\rm Edd}\simeq 1.3\times10^{38}M_{\rm PBH}[\Msun]$ erg/s is the Eddington luminosity, $M_{\rm PBH}$ is the PBH mass, $f(E)$ the probability distribution function for the radiated energy and the dimensionless mass accretion rate $\dot{m}$ is given by 
\begin{equation}\label{eq:mdot}
\dot{m}(z) \simeq 8 \times 10^{-7} \lambda \left(\frac{M_{\rm PBH}}{10 M_\odot}\right)
\left(\frac{n_B(z)}{1\text{cm}^{-3}}\right) \left(\frac{v_{\rm eff}(z)}{10\text{km/s}}\right)^{-3}\,.
\end{equation}
Here $n_{\rm B}(z) = \Omega_B \rho_c (1+z)^3/(\mu m_p)$ is the baryon number density ($\mu\simeq 1.22$ is the mean atomic weight of the neutral gas in units of the proton mass $m_p$), and $v_{\rm eff}$ is the total relative velocity of a PBH with respect to the gas.

In the above we used the convention for the Bondi mass accretion rate given in~\cite{2002apa..book.....F}, see the Eq. (2.36) therein, which for monatomic gas with adiabatic index $\gamma=5/3$ yields a result smaller by a factor of four than the one in Ref.~\cite{Perna:2003ck}. To recast our results in the convention of  Ref.~\cite{Perna:2003ck} it is sufficient to perform the substitution $\lambda\to\lambda'=\lambda/4$.

The spectral emissivity can then be computed as
\begin{equation}\label{eq:emi}
j_E(z) = n_{\rm PBH}(z)\, L_E(z)=f_{\rm PBH} \frac{\Omega_{\rm CDM} \rho_c (1+z)^3}{M_{\rm PBH}}L_E(z)\,,
\end{equation}
where the coefficient $f_{\rm PBH}$ denotes the mass fraction of DM in the form PBHs for a population with a monochromatic mass function centred on $M_{\rm PBH}$.

The effective velocity in Eq.~\eqref{eq:mdot} must account for all the motion components of the accreting PBH relative to the gas. Throughout the cosmic dark ages this comprises two dominant contributions: (i) the relative velocity $v_B(z)$ between PBHs and baryons and (ii) the speed of sound $c_s(z)$ in the baryonic sector. The velocity distribution $v_B$ at this early epoch is well approximated by a Maxwell-Boltzmann distribution. Since $j_E \propto v^{-3(1+\beta)}$, we calculate the average 
\begin{equation}
\langle v^{-3(1+\beta)} |z\rangle = \int \left[v^2 + c_s(z)^2\right]^{-3(1+\beta)/2} f(v|z) \, {\rm d}v,
\end{equation}
where $f(v|z) = \sqrt{2/\pi} v^2 \exp(-v^2/(2\sigma^2(z)))/\sigma^3(z)$.  Here the 1D velocity dispersion is given as $\sigma(z) = 30/\sqrt{3} \text{ km/s} \, (1+z)/(1+z_{\rm rec})$, see for instance Ref.~\cite{Tseliakhovich:2010bj}, $c_s^2(z) = \gamma k_BT_k(z)/(\mu m_p)$ and with $T_k$ being the gas kinetic temperature and $\gamma=5/3$ the adiabatic index. The effective velocity is then defined as $v_{\rm eff}(z) = \langle v^{-3(1+\beta)} |z\rangle^{-1/(3(1+\beta))}$.

In the upper panel of Fig.~\ref{fig:ADAF_template} we show the ADAF spectral templates from Ref.~\cite{2014ARA&A..52..529Y}, which we assume in this study. In the lower panel we display the corresponding cumulative probability distributions for the radiated energy. We find that the most relevant model spectrum for the purposes of the present paper is the lowest one presented in the upper panel, which is characterised by the lowest dimensionless mass accretion rate $\dot{m}$.

Notice that the ADAF spectra of Fig.~\ref{fig:ADAF_template} give $\beta \simeq 0.2$ for the radiative efficiency parameter of Eq.~(\ref{eq:eta}). For the very low accretion rates parameter,  $\beta$ can however be taken as large as $\beta\sim 1$, \eg~\cite{Narayan:2008bv}. In the following, while deriving the constraints for the PBH DM, we therefore allow $\beta$ to vary in the range $[0.2-1]$. As discussed in~\cite{2014ARA&A..52..529Y}, in several situations, for instance when the accretion flow contains non-thermal particles, the prominent inverse Compton bumps of the low $\dot{m}$ models can be smoothed out significantly.

%-------------------------------------------------------------------------------
\subsection{Thermal and ionization history of the baryonic medium} 
\label{sec:history}
%-------------------------------------------------------------------------------

\begin{figure}[t]
	\hskip-.5cm
\includegraphics[width=.9\linewidth]{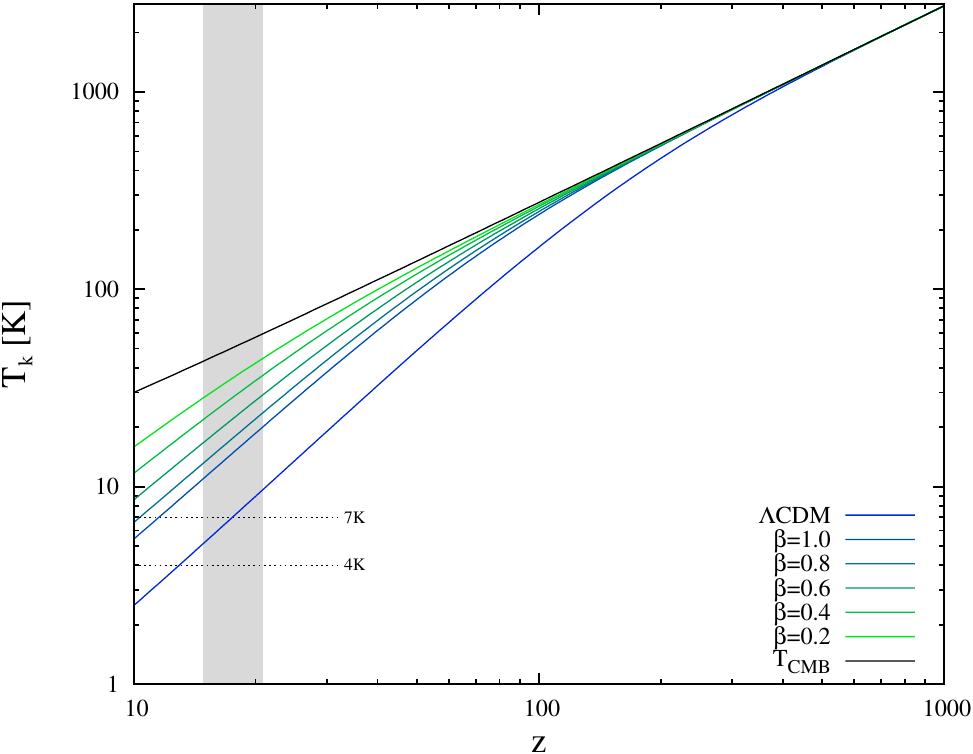}
\caption{Evolution of the gas kinetic temperature, for different values of the radiative efficiency parameter $\beta$ and a mass accretion parameter $\lambda=0.01$. Here $M_{\rm PBH}=10\Msun$ is assumed. The solid black line shows the corresponding evolution of the CMB temperature. The region shaded in grey illustrates the redshift range relevant for the detected EDGES 21-cm absorption feature,  $z\simeq 15-21$. The dotted horizontal lines, marked with the 7K and 4K labels, respectively represent the gas temperature at $z=17.2$ according to the standard $\Lambda$CDM model and the temperature inferred from the EDGES measurement.}
\label{fig:Tk}
\end{figure}

\begin{figure*}
	\centering
\includegraphics[width=0.45\textwidth]{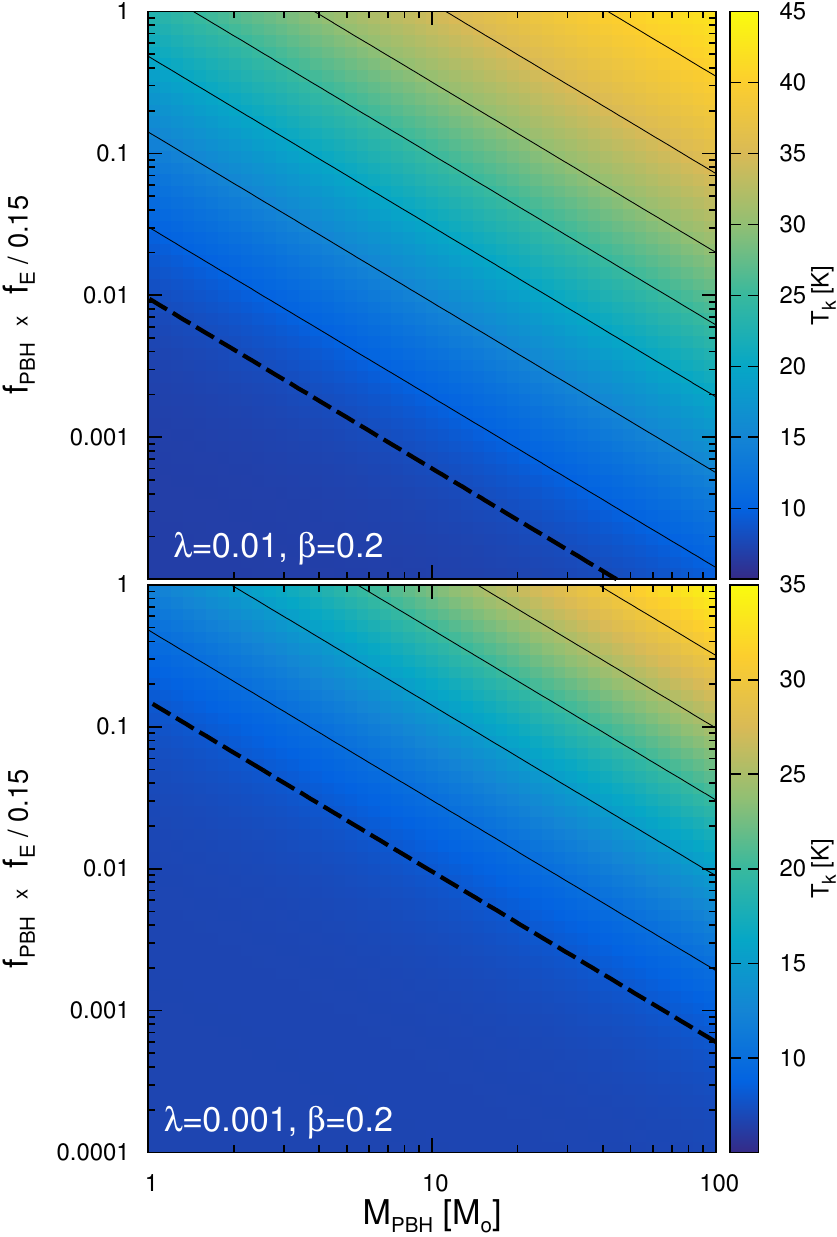} \hspace{5mm}
\includegraphics[width=0.45\textwidth]{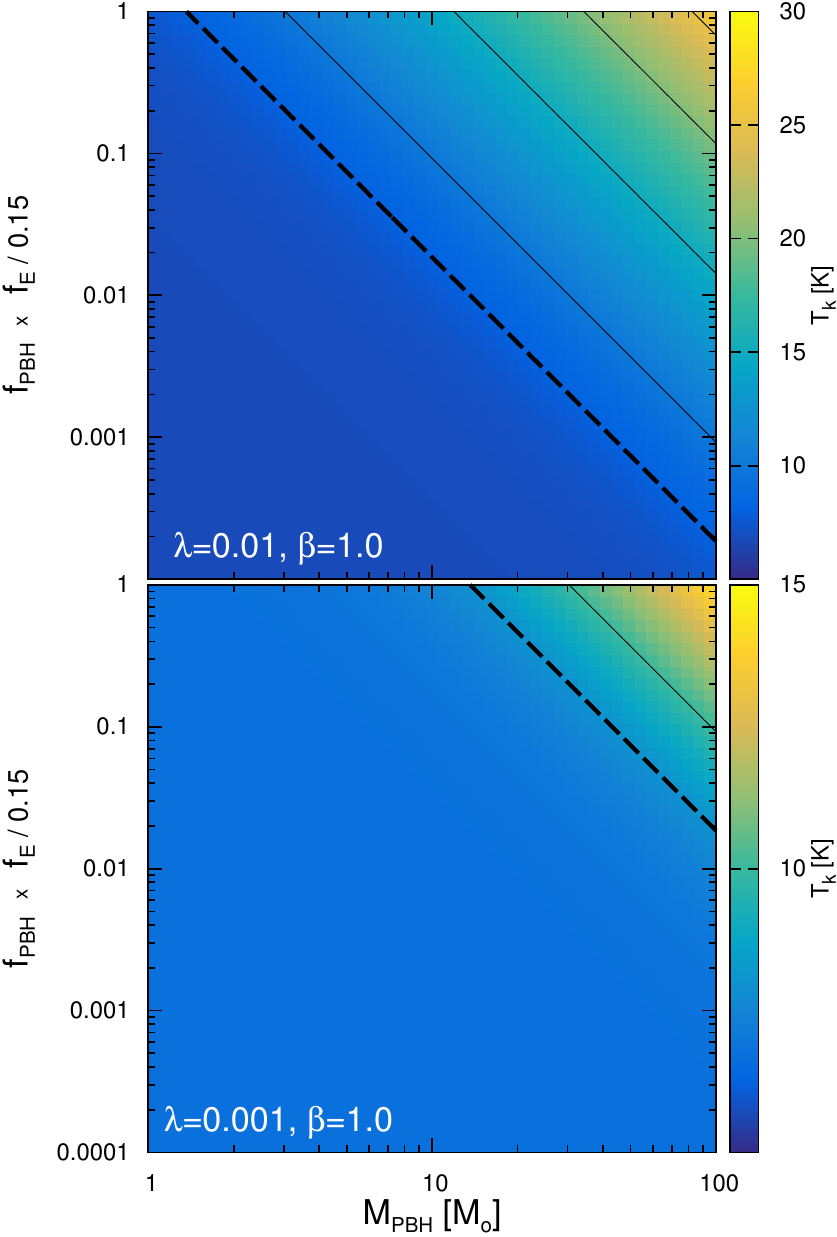}
\caption{Upper bounds on the PBH DM mass fractions $f_{\rm PBH}$ for different maximal gas temperatures (our choice of reference,  $T_k=8$K, is shown with a bold dashed line). The assumed values for the accretion efficiency parameters $\lambda$ and $\beta$ are displayed in the lower left-hand corners of the panels. These calculations assumed that a fraction $f_E$ as large as $15\%$ of the injected energy is efficiently absorbed by the cosmic medium. For different values of this parameter, the $f_{\rm PBH}$ axes should be rescaled by a factor of $f_E/0.15$.}
\label{fig:constraints}
\end{figure*}

Following the Ref.s~\cite{Huetsi:2009ex,Hutsi:2011vx}, we use a modified version of the RECFAST code~\cite{Seager:1999bc} to calculate the thermal and ionization history of the baryonic medium. Since the uncertainties related to the accretion process are by far dominant, we can neglect radiation transfer effects and simply use the `on-spot' approximation to calculate the energy absorption rate of the cosmic medium. We follow in particular the treatment in Ref.~\cite{Padmanabhan:2005es}, with the difference that the function ${\cal F}(z)$, as defined in the Eq.(8) of the reference, does not contain the factor $(1+z)^3$ in our case\footnote{Ref.~\cite{Padmanabhan:2005es} investigated the DM annihilation signal, which scaling with the square of the density is the proportional to $\propto (1+z)^6$}.

The energy absorption rate is then computed as
\begin{equation}
\epsilon(z)=f_E\, \int j_E(z)\, {\rm d}E\,,
\end{equation}
where $f_E < 1$ is the effective energy absorption factor. In the upper panel of Fig.~\ref{fig:mfp_f_param}, we show the photon proper mean free path as observed at redshift $z=17$ for a wide range of photon energies. The dashed horizontal line indicates the horizon size at that epoch. As we can see, for photons with energies below $\sim 1$\,keV, the mean free path is significantly smaller than the horizon size. Therefore, the emitted radiation is efficiently absorbed. 

Thin horizontal dotted lines show instead the mean separation between PBHs with masses $1$, $10$ and $100\Msun$, assuming that all the DM is in this form. As we can see, for ionizing radiation with energies below $\sim 1$\,keV, the mean free path is typically much larger than the average PBH separation. The resulting energy injection is then essentially spatially uniform. 

For larger photon energies, the Universe is instead much more transparent and the `on-spot' approximation starts to break down. For instance, the mean free path photons with  energies $\sim 3.5$\,keV equals the horizon size at $z=17$. As a consequence the `on-spot' approximation has to be corrected. In the lower panel of Fig.~\ref{fig:mfp_f_param} we show the `on-spot' correction factor for the monochromatic photon injections. The relevant calculation details are presented in the Appendix of~\cite{Hektor:2018lec}. In our case the correction factor is averaged over the input ADAF photon spectrum, \ie, $f_E(z)=\int f(E,z)L_E(z)\,{\rm d}E/\int L_E(z)\,{\rm d}E$. We find that at a redshift $z=17$ the correction factor $f_E\simeq 0.12$, whereas for higher redshifts it increases due to the rising opacity of the Universe. Since there are much larger uncertainties in the mass accretion rate, radiative efficiency and in the spectral energy distribution of the emitted light, we decide to use in the following $f_E=0.15$.

In Fig.~\ref{fig:Tk} we show the evolution of the gas temperature as calculated with our modified RECFAST code. Here we set the mass accretion parameter to $\lambda=0.01$, while the radiative efficiency parameter $\beta$ varies in tge range $[0.2-1.0]$. We also show the gas temperature evolution for the standard $\Lambda$CDM model along with the CMB temperature. The Grey shaded region represents the redshift range probed by the EDGES 21-cm absorption feature: $z\simeq 15-21$. The dotted lines carrying a $7$K and $4$K labels mark respectively the gas temperatures at $z=17.2$ according to the $\Lambda$CDM and the EDGES measurement.

\section{Main results}
\label{sec:results}

The best-fit for the 21-cm absorption depth as measured in~\cite{Bowman:2018aa} is $-500$~mK, with a $99\%$ confidence limit corresponding to $-300$~mK. It is then clear that any form of early energy injection will be severely constrained by the EDGES measurement. In particular, in this section we derive a bound on the maximal fraction of PBH DM allowed by requiring that the gas kinetic temperature $T_k$ does not exceed $8$~K, a reference value disfavoured at a confidence level of more than $5\sigma$ by the 21-cm spectrum measurements\footnote{Once a Gaussian distribution is assumed~\cite{Barkana:2018aa}, the adiabatic cooling of the gas in the $\Lambda$CDM leads to a prediction disfavoured at a $\sim 3.8\sigma$ level, barring the contribution of an extra soft photon backgrounds. Accounting for the fact that the distribution is skewed leads to a stronger exclusion.}.

Our model contains in total five free parameters: three accretion specific parameters, $\lambda$, $\beta$ and $f_E$, and two parameters describing monochromatic PBH population, $M_{\rm PBH}$ and $f_{\rm PBH}$. Under our assumptions, the parameters $f_{\rm PBH}$ and $f_E$ are completely degenerate, so that we can only constrain their product. As explained in Sec.~\ref{sec:history}, we use $f_E=0.15$ as our reference value.

We show in Fig.~\ref{fig:constraints} the upper bound on the PBHs DM mass fractions for a range of PBHs masses and for different values of the maximal  gas temperatures allowed $T_k$. Our reference value of $T_k=8$~K is indicated by a bold dashed line. For the plots in the left-hand and right-hand side we respectively assume $\beta=0.2$ and $1.0$. The mass accretion parameter $\lambda$ is taken as $0.01$, and $0.001$ in the top and bottom plots, respectively.

\begin{figure}[t]
\centering
\includegraphics[width=0.45\textwidth]{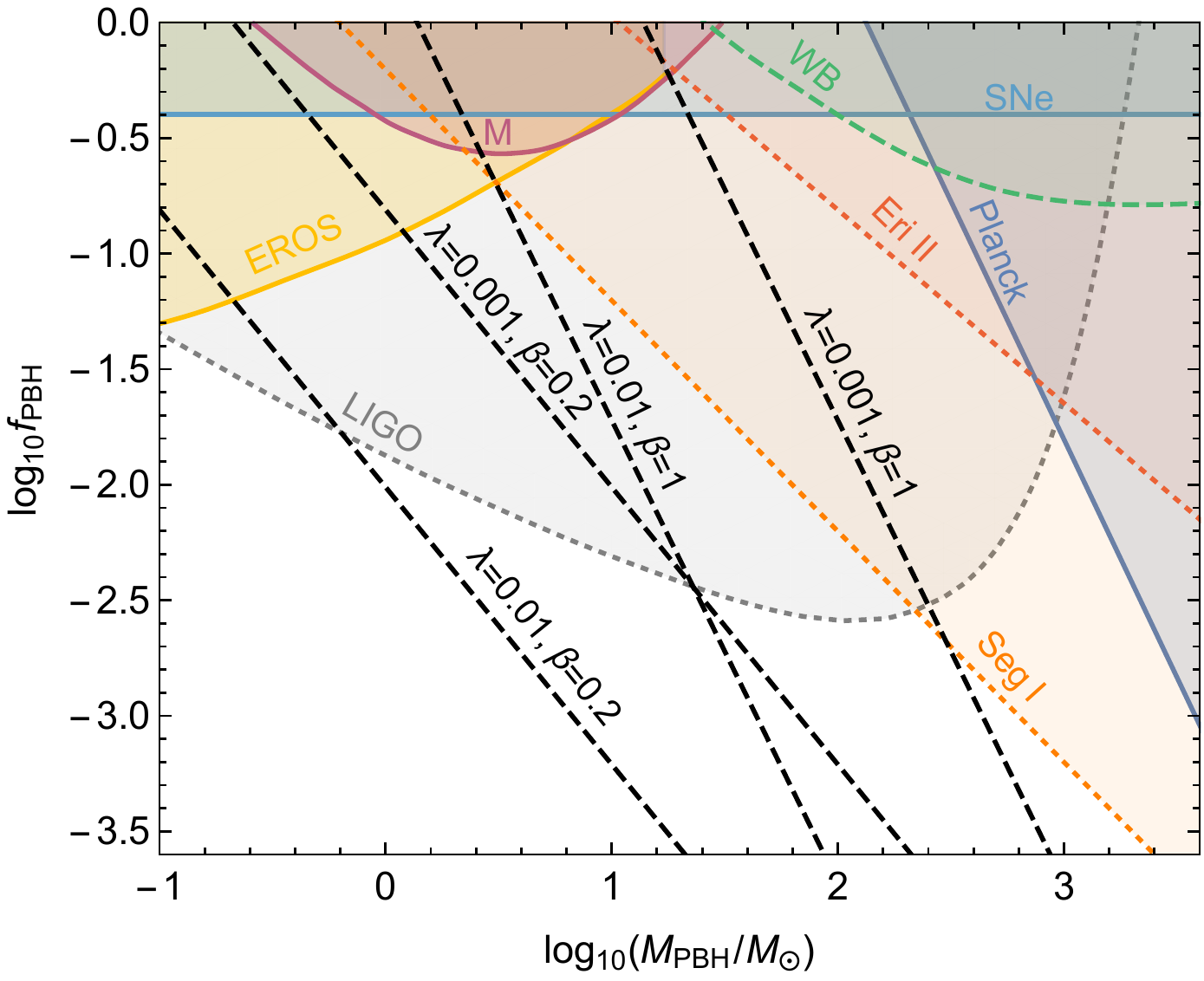}
\caption{Black dashed lines show the constraints implied by the EDGES measurement compared to the bounds on the PBH DM mass function previously considered in the literature. The yellow, purple and light blue regions are excluded by the microlensing results from EROS~\cite{Tisserand:2006zx} and MACHO (M)~\cite{Allsman:2000kg}, and by lack of lensing signatures in type Ia supernovae (SNe)~\cite{Zumalacarregui:2017qqd}, respectively. The dark blue, orange, red and green regions are ruled out by Planck data~\cite{Ali-Haimoud:2016mbv}, survival of stars in Segue I (Seg I)~\cite{Koushiappas:2017chw} and Eridanus II (Eri II)~\cite{Brandt:2016aco}, and the distribution of wide binaries (WB)~\cite{Monroy-Rodriguez:2014ula}, respectively. The gray region (LIGO) is excluded by the non-observation of the stochastic GW background~\cite{Raidal:2017mfl}.}
\label{fig:all_bounds}
\end{figure}

The upper bound that we obtain for $f_{\rm PBH}$ and $T_k=8$K is well approximated as
\begin{equation} \label{eq:bound}
	f_{\rm PBH} \leq C(\beta)\left(\frac{0.15}{f_E}\right)\left(\frac{\lambda}{0.01}\frac{M_{\rm PBH}}{10 M_\odot}\right)^{-(1+\beta)}\,,
\end{equation}
where

\begin{equation*}
C(\beta) =
\begin{cases}
0.00015 + 0.00051\,\beta + 0.0091\,\beta^2 & \text{if $\beta \leq 0.37$} \\
0.019\,\beta^{2.5} & \text{if $\beta > 0.37$}
\end{cases}
\,,
\end{equation*}

and refers to a PBH population characterised by a monochromatic mass function at $M_{\rm PBH}$. The generalisation of the bound to extended PBH mass functions is straightforwardly obtained with the method presented in Ref.~\cite{Carr:2017jsz}.

We show in Fig.~\ref{fig:all_bounds} how the result in Eq.~\eqref{eq:bound} compares to the constraints previously considered in the literature. As we can see, the EDGES measurement implies a bound that is relevant for PBHs populations characterised by a monochromatic mass function above $10\Msun$. We remark that many of the exclusion regions presented in Fig.~\ref{fig:all_bounds} refer to a $2\sigma$ confidence level, whereas our constraint exceeds the $5\sigma$ level. For the constraints previously considered in the literature we show the most conservative bounds. For example, the Planck constraint shown in Fig.~\ref{fig:all_bounds} corresponds to the collisional ionization case in Ref.~\cite{Ali-Haimoud:2016mbv}, which is subdominant with respect to the constraint given, for instrance, in Ref.~\cite{Poulin:2017bwe}, that assumes a different accretion model. As discussed e.g. in Ref.~\cite{Carr:2017jsz}, all of the constraints are subject to various uncertainties. In particular, the LIGO constraint taken from Ref.~\cite{Raidal:2017mfl} assumes the PBH binary formation mechanism introduced in~\cite{Nakamura:1997sm}, which has been criticized e.g. in Ref.~\cite{Clesse:2017bsw}.

In order to account for the impact of an additional soft photon background on Eq.~\eqref{eq:bound}, we computed how the bound scales as a function of the former. We find that the $2\sigma$ upper bound on the gas kinetic temperature $T_k$ is
\begin{equation}
T_k = 4 f_R \,{\rm K} \,,
\end{equation}
where $f_R$ is the soft photon enhancement factor. We remark that any energy injection into the gas is excluded at least at $2\sigma$ confidence level if $f_R<7/4$. For $f_R>7/4$ the correct $2\sigma$ exclusion limits can be obtained by referring to the $T_k$ contours Fig.~\ref{fig:constraints}.

%-------------------------------------------------------------------------------
\section{Conclusions} 
\label{sec:conclusions}
%-------------------------------------------------------------------------------

Motivated by the recent EDGES measurement, we investigated the constraints that the 21-cm absorption feature casts on the maximal fraction of DM in the form of PBHs. In order to account for the energy injection from accreting PBHs, we computed the resulting thermal evolution of the baryonic medium by means of a modified version of the RECFAST code. Due to significant uncertainties in the accretion physics we adopted the `on-spot' approximation to model the energy injections, representing the effects of radiative energy transfer by a single effective parameter. To be as conservative as possible, we also neglected any possible additional sources of energy injection. 

The main results obtained are shown in Fig.~\ref{fig:constraints} and summarised in Fig.~\ref{fig:all_bounds} by means of the bound given in Eq.~\eqref{eq:bound}. With our methodology we find that the EDGES measurement bounds the abundance of PBH populations with masses of ${\cal O}(10) M_\odot$ to be only few percentiles of the measured DM abundance. The 21-cm spectrum therefore constitutes a new and important cosmological observable able to probe the properties of PBHs with a reach comparable to, if not exceeding, that of the constraints previously considered in the literature.  

\noindent
\textbf{Note added:} Simultaneously to our paper, a complementary work~\cite{Clark:2018ghm} appeared, where the EDGES 21-cm observations were used to put constraints on evaporating PBHs.

\section*{Acknowledgements}
We thank the Referee for the constructive suggestions and comments. This work is supported by the grants IUT23-6, IUT26-2, PUT808, PUT799, by EU through the ERDF CoE program grant TK133 and by the Estonian Research Council via the Mobilitas Plus grant MOBTT5.

\appendix

\subsection*{Expected CMB bounds}
In the following we use our parametrized accretion model to estimate the strength of the corresponding PBH DM bounds obtainable from the CMB measurements. We assume that at high redshifts all the energy is efficiently absorbed, \ie\,$f_E=1$, and that the thermal motion of the gas can be neglected in comparison to the DM-baryon streaming velocities. Under these assumptions, the luminosity of accreting PBHs remains constant in time (see Eqs.~\eqref{eq:lumi} and \eqref{eq:mdot}), and the emissivity can thus be given as $j(z)=j_0(1+z)^3$, \ie, similarly to the decaying DM case. For the latter, the CMB bounds have been calculated in Ref.~\cite{Slatyer:2016qyl}, which quantified the $95\%$ confidence level lower bound on the lifetime for DM decaying into photons to be $\tau\sim 10^{24}$\,s. This constraint can be directly converted into a bound relevant to our analysis, yielding

\begin{equation} \label{CMBbound}
f_{\rm PBH}\lesssim \frac{M_{\rm PBH}}{\tau L}\sim\frac{1.4\times10^{-8}\left(\frac{\tau}{10^{24}\,{\rm s}}\right)^{-1}}{100^\beta \left[4.6\times10^{-5}\lambda \left(\frac{M_{\rm PBH}}{10 M_\odot}\right)\right]^{\beta+1}}\,.
\end{equation}

The PBH luminosity of our model agrees reasonably well with the computations of Ref.~\cite{Ali-Haimoud:2016mbv} for $\lambda' = 0.001\, (= 4\lambda)$ and $\beta = 1$. The corresponding bound given by Eq.~\eqref{CMBbound} also matches approximately the CMB bound shown in Fig.~\ref{fig:all_bounds}. 

Once compared to the bound in Eq.~\eqref{eq:bound}, due to the 21-cm measurement, we find that the above CMB constraint is significantly weaker  (their ratio is given as $\sim \frac{1.4\times 10^{-6}}{C(\beta)} \frac{f_E}{0.15}(4.6\times 10^{-5})^{-(\beta+1)}$, for instance $\sim 2.5$ and $\sim 4.5$ orders of magnitude weaker for $\beta=0.2$ and $\beta=1.0$, respectively.

\bibliography{21cm-constraints-on-BHs}

\begin{thebibliography}{65}%
\makeatletter
\providecommand \@ifxundefined [1]{%
 \@ifx{#1\undefined}
}%
\providecommand \@ifnum [1]{%
 \ifnum #1\expandafter \@firstoftwo
 \else \expandafter \@secondoftwo
 \fi
}%
\providecommand \@ifx [1]{%
 \ifx #1\expandafter \@firstoftwo
 \else \expandafter \@secondoftwo
 \fi
}%
\providecommand \natexlab [1]{#1}%
\providecommand \enquote  [1]{``#1''}%
\providecommand \bibnamefont  [1]{#1}%
\providecommand \bibfnamefont [1]{#1}%
\providecommand \citenamefont [1]{#1}%
\providecommand \href@noop [0]{\@secondoftwo}%
\providecommand \href [0]{\begingroup \@sanitize@url \@href}%
\providecommand \@href[1]{\@@startlink{#1}\@@href}%
\providecommand \@@href[1]{\endgroup#1\@@endlink}%
\providecommand \@sanitize@url [0]{\catcode `\\12\catcode `\$12\catcode
  `\&12\catcode `\#12\catcode `\^12\catcode `\_12\catcode `\%12\relax}%
\providecommand \@@startlink[1]{}%
\providecommand \@@endlink[0]{}%
\providecommand \url  [0]{\begingroup\@sanitize@url \@url }%
\providecommand \@url [1]{\endgroup\@href {#1}{\urlprefix }}%
\providecommand \urlprefix  [0]{URL }%
\providecommand \Eprint [0]{\href }%
\providecommand \doibase [0]{http://dx.doi.org/}%
\providecommand \selectlanguage [0]{\@gobble}%
\providecommand \bibinfo  [0]{\@secondoftwo}%
\providecommand \bibfield  [0]{\@secondoftwo}%
\providecommand \translation [1]{[#1]}%
\providecommand \BibitemOpen [0]{}%
\providecommand \bibitemStop [0]{}%
\providecommand \bibitemNoStop [0]{.\EOS\space}%
\providecommand \EOS [0]{\spacefactor3000\relax}%
\providecommand \BibitemShut  [1]{\csname bibitem#1\endcsname}%
\let\auto@bib@innerbib\@empty
%</preamble>
\bibitem [{\citenamefont {Bowman}\ \emph {et~al.}(2018)\citenamefont {Bowman},
  \citenamefont {Rogers}, \citenamefont {Monsalve}, \citenamefont {Mozdzen},\
  and\ \citenamefont {Mahesh}}]{Bowman:2018aa}%
  \BibitemOpen
  \bibfield  {author} {\bibinfo {author} {\bibfnamefont {J.~D.}\ \bibnamefont
  {Bowman}}, \bibinfo {author} {\bibfnamefont {A.~E.~E.}\ \bibnamefont
  {Rogers}}, \bibinfo {author} {\bibfnamefont {R.~A.}\ \bibnamefont
  {Monsalve}}, \bibinfo {author} {\bibfnamefont {T.~J.}\ \bibnamefont
  {Mozdzen}}, \ and\ \bibinfo {author} {\bibfnamefont {N.}~\bibnamefont
  {Mahesh}},\ }\href {http://dx.doi.org/10.1038/nature25792} {\bibfield
  {journal} {\bibinfo  {journal} {Nature}\ }\textbf {\bibinfo {volume} {555}},\
  \bibinfo {pages} {67 EP } (\bibinfo {year} {2018})}\BibitemShut {NoStop}%
\bibitem [{\citenamefont {Barkana}(2018)}]{Barkana:2018aa}%
  \BibitemOpen
  \bibfield  {author} {\bibinfo {author} {\bibfnamefont {R.}~\bibnamefont
  {Barkana}},\ }\href {http://dx.doi.org/10.1038/nature25791} {\bibfield
  {journal} {\bibinfo  {journal} {Nature}\ }\textbf {\bibinfo {volume} {555}},\
  \bibinfo {pages} {71 EP } (\bibinfo {year} {2018})}\BibitemShut {NoStop}%
\bibitem [{\citenamefont {Fialkov}\ \emph {et~al.}(2018)\citenamefont
  {Fialkov}, \citenamefont {Barkana},\ and\ \citenamefont
  {Cohen}}]{Fialkov:2018xre}%
  \BibitemOpen
  \bibfield  {author} {\bibinfo {author} {\bibfnamefont {A.}~\bibnamefont
  {Fialkov}}, \bibinfo {author} {\bibfnamefont {R.}~\bibnamefont {Barkana}}, \
  and\ \bibinfo {author} {\bibfnamefont {A.}~\bibnamefont {Cohen}},\
  }\href@noop {} {\  (\bibinfo {year} {2018})},\ \Eprint
  {http://arxiv.org/abs/1802.10577} {arXiv:1802.10577 [astro-ph.CO]}
  \BibitemShut {NoStop}%
%%CITATION = ARXIV:1802.10577;%%
\bibitem [{\citenamefont {Mu{\~n}oz}\ and\ \citenamefont
  {Loeb}(2018)}]{Munoz:2018pzp}%
  \BibitemOpen
  \bibfield  {author} {\bibinfo {author} {\bibfnamefont {J.~B.}\ \bibnamefont
  {Mu{\~n}oz}}\ and\ \bibinfo {author} {\bibfnamefont {A.}~\bibnamefont
  {Loeb}},\ }\href@noop {} {\  (\bibinfo {year} {2018})},\ \Eprint
  {http://arxiv.org/abs/1802.10094} {arXiv:1802.10094 [astro-ph.CO]}
  \BibitemShut {NoStop}%
%%CITATION = ARXIV:1802.10094;%%
\bibitem [{\citenamefont {Berlin}\ \emph {et~al.}(2018)\citenamefont {Berlin},
  \citenamefont {Hooper}, \citenamefont {Krnjaic},\ and\ \citenamefont
  {McDermott}}]{Berlin:2018sjs}%
  \BibitemOpen
  \bibfield  {author} {\bibinfo {author} {\bibfnamefont {A.}~\bibnamefont
  {Berlin}}, \bibinfo {author} {\bibfnamefont {D.}~\bibnamefont {Hooper}},
  \bibinfo {author} {\bibfnamefont {G.}~\bibnamefont {Krnjaic}}, \ and\
  \bibinfo {author} {\bibfnamefont {S.~D.}\ \bibnamefont {McDermott}},\
  }\href@noop {} {\  (\bibinfo {year} {2018})},\ \Eprint
  {http://arxiv.org/abs/1803.02804} {arXiv:1803.02804 [hep-ph]} \BibitemShut
  {NoStop}%
%%CITATION = ARXIV:1803.02804;%%
\bibitem [{\citenamefont {Barkana}\ \emph {et~al.}(2018)\citenamefont
  {Barkana}, \citenamefont {Outmezguine}, \citenamefont {Redigolo},\ and\
  \citenamefont {Volansky}}]{Barkana:2018qrx}%
  \BibitemOpen
  \bibfield  {author} {\bibinfo {author} {\bibfnamefont {R.}~\bibnamefont
  {Barkana}}, \bibinfo {author} {\bibfnamefont {N.~J.}\ \bibnamefont
  {Outmezguine}}, \bibinfo {author} {\bibfnamefont {D.}~\bibnamefont
  {Redigolo}}, \ and\ \bibinfo {author} {\bibfnamefont {T.}~\bibnamefont
  {Volansky}},\ }\href@noop {} {\  (\bibinfo {year} {2018})},\ \Eprint
  {http://arxiv.org/abs/1803.03091} {arXiv:1803.03091 [hep-ph]} \BibitemShut
  {NoStop}%
%%CITATION = ARXIV:1803.03091;%%
\bibitem [{\citenamefont {Kang}(2018)}]{Kang:2018qhi}%
  \BibitemOpen
  \bibfield  {author} {\bibinfo {author} {\bibfnamefont {Z.}~\bibnamefont
  {Kang}},\ }\href@noop {} {\  (\bibinfo {year} {2018})},\ \Eprint
  {http://arxiv.org/abs/1803.04928} {arXiv:1803.04928 [hep-ph]} \BibitemShut
  {NoStop}%
%%CITATION = ARXIV:1803.04928;%%
\bibitem [{\citenamefont {Dvorkin}\ \emph {et~al.}(2014)\citenamefont
  {Dvorkin}, \citenamefont {Blum},\ and\ \citenamefont
  {Kamionkowski}}]{Dvorkin:2013cea}%
  \BibitemOpen
  \bibfield  {author} {\bibinfo {author} {\bibfnamefont {C.}~\bibnamefont
  {Dvorkin}}, \bibinfo {author} {\bibfnamefont {K.}~\bibnamefont {Blum}}, \
  and\ \bibinfo {author} {\bibfnamefont {M.}~\bibnamefont {Kamionkowski}},\
  }\href {\doibase 10.1103/PhysRevD.89.023519} {\bibfield  {journal} {\bibinfo
  {journal} {Phys. Rev.}\ }\textbf {\bibinfo {volume} {D89}},\ \bibinfo {pages}
  {023519} (\bibinfo {year} {2014})},\ \Eprint {http://arxiv.org/abs/1311.2937}
  {arXiv:1311.2937 [astro-ph.CO]} \BibitemShut {NoStop}%
%%CITATION = ARXIV:1311.2937;%%
\bibitem [{\citenamefont {Tashiro}\ \emph {et~al.}(2014)\citenamefont
  {Tashiro}, \citenamefont {Kadota},\ and\ \citenamefont
  {Silk}}]{Tashiro:2014tsa}%
  \BibitemOpen
  \bibfield  {author} {\bibinfo {author} {\bibfnamefont {H.}~\bibnamefont
  {Tashiro}}, \bibinfo {author} {\bibfnamefont {K.}~\bibnamefont {Kadota}}, \
  and\ \bibinfo {author} {\bibfnamefont {J.}~\bibnamefont {Silk}},\ }\href
  {\doibase 10.1103/PhysRevD.90.083522} {\bibfield  {journal} {\bibinfo
  {journal} {Phys. Rev.}\ }\textbf {\bibinfo {volume} {D90}},\ \bibinfo {pages}
  {083522} (\bibinfo {year} {2014})},\ \Eprint {http://arxiv.org/abs/1408.2571}
  {arXiv:1408.2571 [astro-ph.CO]} \BibitemShut {NoStop}%
%%CITATION = ARXIV:1408.2571;%%
\bibitem [{\citenamefont {Mu{\~n}oz}\ \emph {et~al.}(2015)\citenamefont
  {Mu{\~n}oz}, \citenamefont {Kovetz},\ and\ \citenamefont
  {Ali-Ha{\"i}moud}}]{Munoz:2015bca}%
  \BibitemOpen
  \bibfield  {author} {\bibinfo {author} {\bibfnamefont {J.~B.}\ \bibnamefont
  {Mu{\~n}oz}}, \bibinfo {author} {\bibfnamefont {E.~D.}\ \bibnamefont
  {Kovetz}}, \ and\ \bibinfo {author} {\bibfnamefont {Y.}~\bibnamefont
  {Ali-Ha{\"i}moud}},\ }\href {\doibase 10.1103/PhysRevD.92.083528} {\bibfield
  {journal} {\bibinfo  {journal} {Phys. Rev.}\ }\textbf {\bibinfo {volume}
  {D92}},\ \bibinfo {pages} {083528} (\bibinfo {year} {2015})},\ \Eprint
  {http://arxiv.org/abs/1509.00029} {arXiv:1509.00029 [astro-ph.CO]}
  \BibitemShut {NoStop}%
%%CITATION = ARXIV:1509.00029;%%
\bibitem [{\citenamefont {Mirocha}\ and\ \citenamefont
  {Furlanetto}(2018)}]{Mirocha:2018cih}%
  \BibitemOpen
  \bibfield  {author} {\bibinfo {author} {\bibfnamefont {J.}~\bibnamefont
  {Mirocha}}\ and\ \bibinfo {author} {\bibfnamefont {S.~R.}\ \bibnamefont
  {Furlanetto}},\ }\href@noop {} {\  (\bibinfo {year} {2018})},\ \Eprint
  {http://arxiv.org/abs/1803.03272} {arXiv:1803.03272 [astro-ph.GA]}
  \BibitemShut {NoStop}%
%%CITATION = ARXIV:1803.03272;%%
\bibitem [{\citenamefont {Safarzadeh}\ \emph {et~al.}(2018)\citenamefont
  {Safarzadeh}, \citenamefont {Scannapieco},\ and\ \citenamefont
  {Babul}}]{Safarzadeh:2018hhg}%
  \BibitemOpen
  \bibfield  {author} {\bibinfo {author} {\bibfnamefont {M.}~\bibnamefont
  {Safarzadeh}}, \bibinfo {author} {\bibfnamefont {E.}~\bibnamefont
  {Scannapieco}}, \ and\ \bibinfo {author} {\bibfnamefont {A.}~\bibnamefont
  {Babul}},\ }\href@noop {} {\  (\bibinfo {year} {2018})},\ \Eprint
  {http://arxiv.org/abs/1803.08039} {arXiv:1803.08039 [astro-ph.CO]}
  \BibitemShut {NoStop}%
%%CITATION = ARXIV:1803.08039;%%
\bibitem [{\citenamefont {Hill}\ and\ \citenamefont
  {Baxter}(2018)}]{Hill:2018lfx}%
  \BibitemOpen
  \bibfield  {author} {\bibinfo {author} {\bibfnamefont {J.~C.}\ \bibnamefont
  {Hill}}\ and\ \bibinfo {author} {\bibfnamefont {E.~J.}\ \bibnamefont
  {Baxter}},\ }\href@noop {} {\  (\bibinfo {year} {2018})},\ \Eprint
  {http://arxiv.org/abs/1803.07555} {arXiv:1803.07555 [astro-ph.CO]}
  \BibitemShut {NoStop}%
%%CITATION = ARXIV:1803.07555;%%
\bibitem [{\citenamefont {Costa}\ \emph {et~al.}(2018)\citenamefont {Costa},
  \citenamefont {Landim}, \citenamefont {Wang},\ and\ \citenamefont
  {Abdalla}}]{Costa:2018aoy}%
  \BibitemOpen
  \bibfield  {author} {\bibinfo {author} {\bibfnamefont {A.~A.}\ \bibnamefont
  {Costa}}, \bibinfo {author} {\bibfnamefont {R.~C.~G.}\ \bibnamefont
  {Landim}}, \bibinfo {author} {\bibfnamefont {B.}~\bibnamefont {Wang}}, \ and\
  \bibinfo {author} {\bibfnamefont {E.}~\bibnamefont {Abdalla}},\ }\href@noop
  {} {\  (\bibinfo {year} {2018})},\ \Eprint {http://arxiv.org/abs/1803.06944}
  {arXiv:1803.06944 [astro-ph.CO]} \BibitemShut {NoStop}%
%%CITATION = ARXIV:1803.06944;%%
\bibitem [{\citenamefont {Fraser}\ \emph {et~al.}(2018)\citenamefont {Fraser}
  \emph {et~al.}}]{Fraser:2018acy}%
  \BibitemOpen
  \bibfield  {author} {\bibinfo {author} {\bibfnamefont {S.}~\bibnamefont
  {Fraser}} \emph {et~al.},\ }\href@noop {} {\  (\bibinfo {year} {2018})},\
  \Eprint {http://arxiv.org/abs/1803.03245} {arXiv:1803.03245 [hep-ph]}
  \BibitemShut {NoStop}%
%%CITATION = ARXIV:1803.03245;%%
\bibitem [{\citenamefont {D'Amico}\ \emph {et~al.}(2018)\citenamefont
  {D'Amico}, \citenamefont {Panci},\ and\ \citenamefont
  {Strumia}}]{DAmico:2018sxd}%
  \BibitemOpen
  \bibfield  {author} {\bibinfo {author} {\bibfnamefont {G.}~\bibnamefont
  {D'Amico}}, \bibinfo {author} {\bibfnamefont {P.}~\bibnamefont {Panci}}, \
  and\ \bibinfo {author} {\bibfnamefont {A.}~\bibnamefont {Strumia}},\
  }\href@noop {} {\  (\bibinfo {year} {2018})},\ \Eprint
  {http://arxiv.org/abs/1803.03629} {arXiv:1803.03629 [astro-ph.CO]}
  \BibitemShut {NoStop}%
%%CITATION = ARXIV:1803.03629;%%
\bibitem [{\citenamefont {Yang}\ \emph {et~al.}(2018)\citenamefont {Yang},
  \citenamefont {Huang},\ and\ \citenamefont {Feng}}]{Yang:2018gjd}%
  \BibitemOpen
  \bibfield  {author} {\bibinfo {author} {\bibfnamefont {Y.}~\bibnamefont
  {Yang}}, \bibinfo {author} {\bibfnamefont {X.}~\bibnamefont {Huang}}, \ and\
  \bibinfo {author} {\bibfnamefont {L.}~\bibnamefont {Feng}},\ }\href@noop {}
  {\  (\bibinfo {year} {2018})},\ \Eprint {http://arxiv.org/abs/1803.05803}
  {arXiv:1803.05803 [astro-ph.CO]} \BibitemShut {NoStop}%
%%CITATION = ARXIV:1803.05803;%%
\bibitem [{\citenamefont {Ewall-Wice}\ \emph {et~al.}(2018)\citenamefont
  {Ewall-Wice}, \citenamefont {Chang}, \citenamefont {Lazio}, \citenamefont
  {Dore}, \citenamefont {Seiffert},\ and\ \citenamefont
  {Monsalve}}]{Ewall-Wice:2018bzf}%
  \BibitemOpen
  \bibfield  {author} {\bibinfo {author} {\bibfnamefont {A.}~\bibnamefont
  {Ewall-Wice}}, \bibinfo {author} {\bibfnamefont {T.~C.}\ \bibnamefont
  {Chang}}, \bibinfo {author} {\bibfnamefont {J.}~\bibnamefont {Lazio}},
  \bibinfo {author} {\bibfnamefont {O.}~\bibnamefont {Dore}}, \bibinfo {author}
  {\bibfnamefont {M.}~\bibnamefont {Seiffert}}, \ and\ \bibinfo {author}
  {\bibfnamefont {R.~A.}\ \bibnamefont {Monsalve}},\ }\href@noop {} {\
  (\bibinfo {year} {2018})},\ \Eprint {http://arxiv.org/abs/1803.01815}
  {arXiv:1803.01815 [astro-ph.CO]} \BibitemShut {NoStop}%
%%CITATION = ARXIV:1803.01815;%%
\bibitem [{\citenamefont {Pospelov}\ \emph {et~al.}(2018)\citenamefont
  {Pospelov}, \citenamefont {Pradler}, \citenamefont {Ruderman},\ and\
  \citenamefont {Urbano}}]{Pospelov:2018kdh}%
  \BibitemOpen
  \bibfield  {author} {\bibinfo {author} {\bibfnamefont {M.}~\bibnamefont
  {Pospelov}}, \bibinfo {author} {\bibfnamefont {J.}~\bibnamefont {Pradler}},
  \bibinfo {author} {\bibfnamefont {J.~T.}\ \bibnamefont {Ruderman}}, \ and\
  \bibinfo {author} {\bibfnamefont {A.}~\bibnamefont {Urbano}},\ }\href@noop {}
  {\  (\bibinfo {year} {2018})},\ \Eprint {http://arxiv.org/abs/1803.07048}
  {arXiv:1803.07048 [hep-ph]} \BibitemShut {NoStop}%
%%CITATION = ARXIV:1803.07048;%%
\bibitem [{\citenamefont {{Fixsen}}\ \emph {et~al.}(2011)\citenamefont
  {{Fixsen}}, \citenamefont {{Kogut}}, \citenamefont {{Levin}}, \citenamefont
  {{Limon}}, \citenamefont {{Lubin}}, \citenamefont {{Mirel}}, \citenamefont
  {{Seiffert}}, \citenamefont {{Singal}}, \citenamefont {{Wollack}},
  \citenamefont {{Villela}},\ and\ \citenamefont
  {{Wuensche}}}]{2011ApJ...734....5F}%
  \BibitemOpen
  \bibfield  {author} {\bibinfo {author} {\bibfnamefont {D.~J.}\ \bibnamefont
  {{Fixsen}}}, \bibinfo {author} {\bibfnamefont {A.}~\bibnamefont {{Kogut}}},
  \bibinfo {author} {\bibfnamefont {S.}~\bibnamefont {{Levin}}}, \bibinfo
  {author} {\bibfnamefont {M.}~\bibnamefont {{Limon}}}, \bibinfo {author}
  {\bibfnamefont {P.}~\bibnamefont {{Lubin}}}, \bibinfo {author} {\bibfnamefont
  {P.}~\bibnamefont {{Mirel}}}, \bibinfo {author} {\bibfnamefont
  {M.}~\bibnamefont {{Seiffert}}}, \bibinfo {author} {\bibfnamefont
  {J.}~\bibnamefont {{Singal}}}, \bibinfo {author} {\bibfnamefont
  {E.}~\bibnamefont {{Wollack}}}, \bibinfo {author} {\bibfnamefont
  {T.}~\bibnamefont {{Villela}}}, \ and\ \bibinfo {author} {\bibfnamefont
  {C.~A.}\ \bibnamefont {{Wuensche}}},\ }\href {\doibase
  10.1088/0004-637X/734/1/5} {\bibfield  {journal} {\bibinfo  {journal} {\apj}\
  }\textbf {\bibinfo {volume} {734}},\ \bibinfo {eid} {5} (\bibinfo {year}
  {2011})},\ \Eprint {http://arxiv.org/abs/0901.0555} {arXiv:0901.0555}
  \BibitemShut {NoStop}%
\bibitem [{\citenamefont {Feng}\ and\ \citenamefont
  {Holder}(2018)}]{Feng:2018rje}%
  \BibitemOpen
  \bibfield  {author} {\bibinfo {author} {\bibfnamefont {C.}~\bibnamefont
  {Feng}}\ and\ \bibinfo {author} {\bibfnamefont {G.}~\bibnamefont {Holder}},\
  }\href@noop {} {\  (\bibinfo {year} {2018})},\ \Eprint
  {http://arxiv.org/abs/1802.07432} {arXiv:1802.07432 [astro-ph.CO]}
  \BibitemShut {NoStop}%
%%CITATION = ARXIV:1802.07432;%%
\bibitem [{\citenamefont {Hawking}(1971)}]{Hawking:1971ei}%
  \BibitemOpen
  \bibfield  {author} {\bibinfo {author} {\bibfnamefont {S.}~\bibnamefont
  {Hawking}},\ }\href@noop {} {\bibfield  {journal} {\bibinfo  {journal} {Mon.
  Not. Roy. Astron. Soc.}\ }\textbf {\bibinfo {volume} {152}},\ \bibinfo
  {pages} {75} (\bibinfo {year} {1971})}\BibitemShut {NoStop}%
%%CITATION = MNRAA,152,75;%%
\bibitem [{\citenamefont {Carr}\ and\ \citenamefont
  {Hawking}(1974)}]{Carr:1974nx}%
  \BibitemOpen
  \bibfield  {author} {\bibinfo {author} {\bibfnamefont {B.~J.}\ \bibnamefont
  {Carr}}\ and\ \bibinfo {author} {\bibfnamefont {S.~W.}\ \bibnamefont
  {Hawking}},\ }\href@noop {} {\bibfield  {journal} {\bibinfo  {journal} {Mon.
  Not. Roy. Astron. Soc.}\ }\textbf {\bibinfo {volume} {168}},\ \bibinfo
  {pages} {399} (\bibinfo {year} {1974})}\BibitemShut {NoStop}%
%%CITATION = MNRAA,168,399;%%
\bibitem [{\citenamefont {Carr}(1975)}]{Carr:1975qj}%
  \BibitemOpen
  \bibfield  {author} {\bibinfo {author} {\bibfnamefont {B.~J.}\ \bibnamefont
  {Carr}},\ }\href {\doibase 10.1086/153853} {\bibfield  {journal} {\bibinfo
  {journal} {Astrophys. J.}\ }\textbf {\bibinfo {volume} {201}},\ \bibinfo
  {pages} {1} (\bibinfo {year} {1975})}\BibitemShut {NoStop}%
%%CITATION = ASJOA,201,1;%%
\bibitem [{\citenamefont {{Meszaros}}(1975)}]{1975A&A....38....5M}%
  \BibitemOpen
  \bibfield  {author} {\bibinfo {author} {\bibfnamefont {P.}~\bibnamefont
  {{Meszaros}}},\ }\href@noop {} {\bibfield  {journal} {\bibinfo  {journal}
  {Astron. Astrophys.}\ }\textbf {\bibinfo {volume} {38}},\ \bibinfo {pages}
  {5} (\bibinfo {year} {1975})}\BibitemShut {NoStop}%
\bibitem [{\citenamefont {Chapline}(1975)}]{CHAPLINE:1975aa}%
  \BibitemOpen
  \bibfield  {author} {\bibinfo {author} {\bibfnamefont {G.~F.}\ \bibnamefont
  {Chapline}},\ }\href {http://dx.doi.org/10.1038/253251a0} {\bibfield
  {journal} {\bibinfo  {journal} {Nature}\ }\textbf {\bibinfo {volume} {253}},\
  \bibinfo {pages} {251} (\bibinfo {year} {1975})}\BibitemShut {NoStop}%
\bibitem [{\citenamefont {Abbott}\ \emph
  {et~al.}(2016{\natexlab{a}})\citenamefont {Abbott} \emph
  {et~al.}}]{Abbott:2016blz}%
  \BibitemOpen
  \bibfield  {author} {\bibinfo {author} {\bibfnamefont {B.~P.}\ \bibnamefont
  {Abbott}} \emph {et~al.} (\bibinfo {collaboration} {Virgo, LIGO
  Scientific}),\ }\href {\doibase 10.1103/PhysRevLett.116.061102} {\bibfield
  {journal} {\bibinfo  {journal} {Phys. Rev. Lett.}\ }\textbf {\bibinfo
  {volume} {116}},\ \bibinfo {pages} {061102} (\bibinfo {year}
  {2016}{\natexlab{a}})},\ \Eprint {http://arxiv.org/abs/1602.03837}
  {arXiv:1602.03837 [gr-qc]} \BibitemShut {NoStop}%
%%CITATION = ARXIV:1602.03837;%%
\bibitem [{\citenamefont {Abbott}\ \emph
  {et~al.}(2016{\natexlab{b}})\citenamefont {Abbott} \emph
  {et~al.}}]{Abbott:2016nmj}%
  \BibitemOpen
  \bibfield  {author} {\bibinfo {author} {\bibfnamefont {B.~P.}\ \bibnamefont
  {Abbott}} \emph {et~al.} (\bibinfo {collaboration} {Virgo, LIGO
  Scientific}),\ }\href {\doibase 10.1103/PhysRevLett.116.241103} {\bibfield
  {journal} {\bibinfo  {journal} {Phys. Rev. Lett.}\ }\textbf {\bibinfo
  {volume} {116}},\ \bibinfo {pages} {241103} (\bibinfo {year}
  {2016}{\natexlab{b}})},\ \Eprint {http://arxiv.org/abs/1606.04855}
  {arXiv:1606.04855 [gr-qc]} \BibitemShut {NoStop}%
%%CITATION = ARXIV:1606.04855;%%
\bibitem [{\citenamefont {Carr}\ \emph {et~al.}(2016)\citenamefont {Carr},
  \citenamefont {Kuhnel},\ and\ \citenamefont {Sandstad}}]{Carr:2016drx}%
  \BibitemOpen
  \bibfield  {author} {\bibinfo {author} {\bibfnamefont {B.}~\bibnamefont
  {Carr}}, \bibinfo {author} {\bibfnamefont {F.}~\bibnamefont {Kuhnel}}, \ and\
  \bibinfo {author} {\bibfnamefont {M.}~\bibnamefont {Sandstad}},\ }\href
  {\doibase 10.1103/PhysRevD.94.083504} {\bibfield  {journal} {\bibinfo
  {journal} {Phys. Rev.}\ }\textbf {\bibinfo {volume} {D94}},\ \bibinfo {pages}
  {083504} (\bibinfo {year} {2016})},\ \Eprint
  {http://arxiv.org/abs/1607.06077} {arXiv:1607.06077 [astro-ph.CO]}
  \BibitemShut {NoStop}%
%%CITATION = ARXIV:1607.06077;%%
\bibitem [{\citenamefont {Kashlinsky}(2016)}]{Kashlinsky:2016sdv}%
  \BibitemOpen
  \bibfield  {author} {\bibinfo {author} {\bibfnamefont {A.}~\bibnamefont
  {Kashlinsky}},\ }\href {\doibase 10.3847/2041-8205/823/2/L25} {\bibfield
  {journal} {\bibinfo  {journal} {Astrophys. J.}\ }\textbf {\bibinfo {volume}
  {823}},\ \bibinfo {pages} {L25} (\bibinfo {year} {2016})},\ \Eprint
  {http://arxiv.org/abs/1605.04023} {arXiv:1605.04023 [astro-ph.CO]}
  \BibitemShut {NoStop}%
%%CITATION = ARXIV:1605.04023;%%
\bibitem [{\citenamefont {Bird}\ \emph {et~al.}(2016)\citenamefont {Bird},
  \citenamefont {Cholis}, \citenamefont {Mu{\~n}oz}, \citenamefont
  {Ali-Ha{\"\i}moud}, \citenamefont {Kamionkowski}, \citenamefont {Kovetz},
  \citenamefont {Raccanelli},\ and\ \citenamefont {Riess}}]{Bird:2016dcv}%
  \BibitemOpen
  \bibfield  {author} {\bibinfo {author} {\bibfnamefont {S.}~\bibnamefont
  {Bird}}, \bibinfo {author} {\bibfnamefont {I.}~\bibnamefont {Cholis}},
  \bibinfo {author} {\bibfnamefont {J.~B.}\ \bibnamefont {Mu{\~n}oz}}, \bibinfo
  {author} {\bibfnamefont {Y.}~\bibnamefont {Ali-Ha{\"\i}moud}}, \bibinfo
  {author} {\bibfnamefont {M.}~\bibnamefont {Kamionkowski}}, \bibinfo {author}
  {\bibfnamefont {E.~D.}\ \bibnamefont {Kovetz}}, \bibinfo {author}
  {\bibfnamefont {A.}~\bibnamefont {Raccanelli}}, \ and\ \bibinfo {author}
  {\bibfnamefont {A.~G.}\ \bibnamefont {Riess}},\ }\href {\doibase
  10.1103/PhysRevLett.116.201301} {\bibfield  {journal} {\bibinfo  {journal}
  {Phys. Rev. Lett.}\ }\textbf {\bibinfo {volume} {116}},\ \bibinfo {pages}
  {201301} (\bibinfo {year} {2016})},\ \Eprint
  {http://arxiv.org/abs/1603.00464} {arXiv:1603.00464 [astro-ph.CO]}
  \BibitemShut {NoStop}%
%%CITATION = ARXIV:1603.00464;%%
\bibitem [{\citenamefont {Clesse}\ and\ \citenamefont
  {Garc{\'\i}a-Bellido}(2016)}]{Clesse:2016vqa}%
  \BibitemOpen
  \bibfield  {author} {\bibinfo {author} {\bibfnamefont {S.}~\bibnamefont
  {Clesse}}\ and\ \bibinfo {author} {\bibfnamefont {J.}~\bibnamefont
  {Garc{\'\i}a-Bellido}},\ }\href {\doibase 10.1016/j.dark.2016.10.002}
  {\bibfield  {journal} {\bibinfo  {journal} {Phys. Dark Univ.}\ }\textbf
  {\bibinfo {volume} {10}},\ \bibinfo {pages} {002} (\bibinfo {year} {2016})},\
  \Eprint {http://arxiv.org/abs/1603.05234} {arXiv:1603.05234 [astro-ph.CO]}
  \BibitemShut {NoStop}%
%%CITATION = ARXIV:1603.05234;%%
\bibitem [{\citenamefont {Sasaki}\ \emph {et~al.}(2016)\citenamefont {Sasaki},
  \citenamefont {Suyama}, \citenamefont {Tanaka},\ and\ \citenamefont
  {Yokoyama}}]{Sasaki:2016jop}%
  \BibitemOpen
  \bibfield  {author} {\bibinfo {author} {\bibfnamefont {M.}~\bibnamefont
  {Sasaki}}, \bibinfo {author} {\bibfnamefont {T.}~\bibnamefont {Suyama}},
  \bibinfo {author} {\bibfnamefont {T.}~\bibnamefont {Tanaka}}, \ and\ \bibinfo
  {author} {\bibfnamefont {S.}~\bibnamefont {Yokoyama}},\ }\href {\doibase
  10.1103/PhysRevLett.117.061101} {\bibfield  {journal} {\bibinfo  {journal}
  {Phys. Rev. Lett.}\ }\textbf {\bibinfo {volume} {117}},\ \bibinfo {pages}
  {061101} (\bibinfo {year} {2016})},\ \Eprint
  {http://arxiv.org/abs/1603.08338} {arXiv:1603.08338 [astro-ph.CO]}
  \BibitemShut {NoStop}%
%%CITATION = ARXIV:1603.08338;%%
\bibitem [{\citenamefont {Sasaki}\ \emph {et~al.}(2018)\citenamefont {Sasaki},
  \citenamefont {Suyama}, \citenamefont {Tanaka},\ and\ \citenamefont
  {Yokoyama}}]{Sasaki:2018dmp}%
  \BibitemOpen
  \bibfield  {author} {\bibinfo {author} {\bibfnamefont {M.}~\bibnamefont
  {Sasaki}}, \bibinfo {author} {\bibfnamefont {T.}~\bibnamefont {Suyama}},
  \bibinfo {author} {\bibfnamefont {T.}~\bibnamefont {Tanaka}}, \ and\ \bibinfo
  {author} {\bibfnamefont {S.}~\bibnamefont {Yokoyama}},\ }\href {\doibase
  10.1088/1361-6382/aaa7b4} {\bibfield  {journal} {\bibinfo  {journal} {Class.
  Quant. Grav.}\ }\textbf {\bibinfo {volume} {35}},\ \bibinfo {pages} {063001}
  (\bibinfo {year} {2018})},\ \Eprint {http://arxiv.org/abs/1801.05235}
  {arXiv:1801.05235 [astro-ph.CO]} \BibitemShut {NoStop}%
%%CITATION = ARXIV:1801.05235;%%
\bibitem [{\citenamefont {Carr}\ \emph {et~al.}(2017)\citenamefont {Carr},
  \citenamefont {Raidal}, \citenamefont {Tenkanen}, \citenamefont {Vaskonen},\
  and\ \citenamefont {Veerm{\"a}e}}]{Carr:2017jsz}%
  \BibitemOpen
  \bibfield  {author} {\bibinfo {author} {\bibfnamefont {B.}~\bibnamefont
  {Carr}}, \bibinfo {author} {\bibfnamefont {M.}~\bibnamefont {Raidal}},
  \bibinfo {author} {\bibfnamefont {T.}~\bibnamefont {Tenkanen}}, \bibinfo
  {author} {\bibfnamefont {V.}~\bibnamefont {Vaskonen}}, \ and\ \bibinfo
  {author} {\bibfnamefont {H.}~\bibnamefont {Veerm{\"a}e}},\ }\href {\doibase
  10.1103/PhysRevD.96.023514} {\bibfield  {journal} {\bibinfo  {journal} {Phys.
  Rev.}\ }\textbf {\bibinfo {volume} {D96}},\ \bibinfo {pages} {023514}
  (\bibinfo {year} {2017})},\ \Eprint {http://arxiv.org/abs/1705.05567}
  {arXiv:1705.05567 [astro-ph.CO]} \BibitemShut {NoStop}%
%%CITATION = ARXIV:1705.05567;%%
\bibitem [{\citenamefont {Raidal}\ \emph {et~al.}(2018)\citenamefont {Raidal},
  \citenamefont {Solodukhin}, \citenamefont {Vaskonen},\ and\ \citenamefont
  {Veerm{\"a}e}}]{Raidal:2018eoo}%
  \BibitemOpen
  \bibfield  {author} {\bibinfo {author} {\bibfnamefont {M.}~\bibnamefont
  {Raidal}}, \bibinfo {author} {\bibfnamefont {S.}~\bibnamefont {Solodukhin}},
  \bibinfo {author} {\bibfnamefont {V.}~\bibnamefont {Vaskonen}}, \ and\
  \bibinfo {author} {\bibfnamefont {H.}~\bibnamefont {Veerm{\"a}e}},\
  }\href@noop {} {\  (\bibinfo {year} {2018})},\ \Eprint
  {http://arxiv.org/abs/1802.07728} {arXiv:1802.07728 [astro-ph.CO]}
  \BibitemShut {NoStop}%
%%CITATION = ARXIV:1802.07728;%%
\bibitem [{\citenamefont {Garc{\'\i}a-Bellido}\ and\ \citenamefont
  {Clesse}(2018)}]{Garcia-Bellido:2017xvr}%
  \BibitemOpen
  \bibfield  {author} {\bibinfo {author} {\bibfnamefont {J.}~\bibnamefont
  {Garc{\'\i}a-Bellido}}\ and\ \bibinfo {author} {\bibfnamefont
  {S.}~\bibnamefont {Clesse}},\ }\href {\doibase 10.1016/j.dark.2018.01.001}
  {\bibfield  {journal} {\bibinfo  {journal} {Phys. Dark Univ.}\ }\textbf
  {\bibinfo {volume} {19}},\ \bibinfo {pages} {144} (\bibinfo {year} {2018})},\
  \Eprint {http://arxiv.org/abs/1710.04694} {arXiv:1710.04694 [astro-ph.CO]}
  \BibitemShut {NoStop}%
%%CITATION = ARXIV:1710.04694;%%
\bibitem [{\citenamefont {Clesse}\ and\ \citenamefont
  {Garc{\'\i}a-Bellido}(2017)}]{Clesse:2017bsw}%
  \BibitemOpen
  \bibfield  {author} {\bibinfo {author} {\bibfnamefont {S.}~\bibnamefont
  {Clesse}}\ and\ \bibinfo {author} {\bibfnamefont {J.}~\bibnamefont
  {Garc{\'\i}a-Bellido}},\ }\href@noop {} {\  (\bibinfo {year} {2017})},\
  \Eprint {http://arxiv.org/abs/1711.10458} {arXiv:1711.10458 [astro-ph.CO]}
  \BibitemShut {NoStop}%
%%CITATION = ARXIV:1711.10458;%%
\bibitem [{\citenamefont {Garcia-Bellido}\ \emph {et~al.}(2017)\citenamefont
  {Garcia-Bellido}, \citenamefont {Clesse},\ and\ \citenamefont
  {Fleury}}]{Garcia-Bellido:2017imq}%
  \BibitemOpen
  \bibfield  {author} {\bibinfo {author} {\bibfnamefont {J.}~\bibnamefont
  {Garcia-Bellido}}, \bibinfo {author} {\bibfnamefont {S.}~\bibnamefont
  {Clesse}}, \ and\ \bibinfo {author} {\bibfnamefont {P.}~\bibnamefont
  {Fleury}},\ }\href@noop {} {\  (\bibinfo {year} {2017})},\ \Eprint
  {http://arxiv.org/abs/1712.06574} {arXiv:1712.06574 [astro-ph.CO]}
  \BibitemShut {NoStop}%
%%CITATION = ARXIV:1712.06574;%%
\bibitem [{\citenamefont {Ricotti}\ \emph {et~al.}(2008)\citenamefont
  {Ricotti}, \citenamefont {Ostriker},\ and\ \citenamefont
  {Mack}}]{Ricotti:2007au}%
  \BibitemOpen
  \bibfield  {author} {\bibinfo {author} {\bibfnamefont {M.}~\bibnamefont
  {Ricotti}}, \bibinfo {author} {\bibfnamefont {J.~P.}\ \bibnamefont
  {Ostriker}}, \ and\ \bibinfo {author} {\bibfnamefont {K.~J.}\ \bibnamefont
  {Mack}},\ }\href {\doibase 10.1086/587831} {\bibfield  {journal} {\bibinfo
  {journal} {Astrophys. J.}\ }\textbf {\bibinfo {volume} {680}},\ \bibinfo
  {pages} {829} (\bibinfo {year} {2008})},\ \Eprint
  {http://arxiv.org/abs/0709.0524} {arXiv:0709.0524 [astro-ph]} \BibitemShut
  {NoStop}%
%%CITATION = ARXIV:0709.0524;%%
\bibitem [{\citenamefont {Horowitz}(2016)}]{Horowitz:2016lib}%
  \BibitemOpen
  \bibfield  {author} {\bibinfo {author} {\bibfnamefont {B.}~\bibnamefont
  {Horowitz}},\ }\href@noop {} {\  (\bibinfo {year} {2016})},\ \Eprint
  {http://arxiv.org/abs/1612.07264} {arXiv:1612.07264 [astro-ph.CO]}
  \BibitemShut {NoStop}%
%%CITATION = ARXIV:1612.07264;%%
\bibitem [{\citenamefont {{Ali-Ha\"imoud}}\ and\ \citenamefont
  {Kamionkowski}(2017)}]{Ali-Haimoud:2016mbv}%
  \BibitemOpen
  \bibfield  {author} {\bibinfo {author} {\bibfnamefont {Y.}~\bibnamefont
  {{Ali-Ha\"imoud}}}\ and\ \bibinfo {author} {\bibfnamefont {M.}~\bibnamefont
  {Kamionkowski}},\ }\href {\doibase 10.1103/PhysRevD.95.043534} {\bibfield
  {journal} {\bibinfo  {journal} {Phys. Rev.}\ }\textbf {\bibinfo {volume}
  {D95}},\ \bibinfo {pages} {043534} (\bibinfo {year} {2017})},\ \Eprint
  {http://arxiv.org/abs/1612.05644} {arXiv:1612.05644 [astro-ph.CO]}
  \BibitemShut {NoStop}%
%%CITATION = ARXIV:1612.05644;%%
\bibitem [{\citenamefont {Poulin}\ \emph {et~al.}(2017)\citenamefont {Poulin},
  \citenamefont {Serpico}, \citenamefont {Calore}, \citenamefont {Clesse},\
  and\ \citenamefont {Kohri}}]{Poulin:2017bwe}%
  \BibitemOpen
  \bibfield  {author} {\bibinfo {author} {\bibfnamefont {V.}~\bibnamefont
  {Poulin}}, \bibinfo {author} {\bibfnamefont {P.~D.}\ \bibnamefont {Serpico}},
  \bibinfo {author} {\bibfnamefont {F.}~\bibnamefont {Calore}}, \bibinfo
  {author} {\bibfnamefont {S.}~\bibnamefont {Clesse}}, \ and\ \bibinfo {author}
  {\bibfnamefont {K.}~\bibnamefont {Kohri}},\ }\href {\doibase
  10.1103/PhysRevD.96.083524} {\bibfield  {journal} {\bibinfo  {journal} {Phys.
  Rev.}\ }\textbf {\bibinfo {volume} {D96}},\ \bibinfo {pages} {083524}
  (\bibinfo {year} {2017})},\ \Eprint {http://arxiv.org/abs/1707.04206}
  {arXiv:1707.04206 [astro-ph.CO]} \BibitemShut {NoStop}%
%%CITATION = ARXIV:1707.04206;%%
\bibitem [{\citenamefont {{Yuan}}\ and\ \citenamefont
  {{Narayan}}(2014)}]{2014ARA&A..52..529Y}%
  \BibitemOpen
  \bibfield  {author} {\bibinfo {author} {\bibfnamefont {F.}~\bibnamefont
  {{Yuan}}}\ and\ \bibinfo {author} {\bibfnamefont {R.}~\bibnamefont
  {{Narayan}}},\ }\href {\doibase 10.1146/annurev-astro-082812-141003}
  {\bibfield  {journal} {\bibinfo  {journal} {Annual Review of Astronomy and
  Astrophysics}\ }\textbf {\bibinfo {volume} {52}},\ \bibinfo {pages} {529}
  (\bibinfo {year} {2014})},\ \Eprint {http://arxiv.org/abs/1401.0586}
  {arXiv:1401.0586 [astro-ph.HE]} \BibitemShut {NoStop}%
\bibitem [{\citenamefont {Bondi}(1952)}]{Bondi:1952ni}%
  \BibitemOpen
  \bibfield  {author} {\bibinfo {author} {\bibfnamefont {H.}~\bibnamefont
  {Bondi}},\ }\href@noop {} {\bibfield  {journal} {\bibinfo  {journal} {Mon.
  Not. Roy. Astron. Soc.}\ }\textbf {\bibinfo {volume} {112}},\ \bibinfo
  {pages} {195} (\bibinfo {year} {1952})}\BibitemShut {NoStop}%
%%CITATION = MNRAA,112,195;%%
\bibitem [{\citenamefont {{Frank}}\ \emph {et~al.}(2002)\citenamefont
  {{Frank}}, \citenamefont {{King}},\ and\ \citenamefont
  {{Raine}}}]{2002apa..book.....F}%
  \BibitemOpen
  \bibfield  {author} {\bibinfo {author} {\bibfnamefont {J.}~\bibnamefont
  {{Frank}}}, \bibinfo {author} {\bibfnamefont {A.}~\bibnamefont {{King}}}, \
  and\ \bibinfo {author} {\bibfnamefont {D.~J.}\ \bibnamefont {{Raine}}},\
  }\href@noop {} {\emph {\bibinfo {title} {{Accretion Power in Astrophysics:
  Third Edition}}}}\ (\bibinfo  {publisher} {Cambridge University Press},\
  \bibinfo {year} {2002})\ p.\ \bibinfo {pages} {398}\BibitemShut {NoStop}%
\bibitem [{\citenamefont {Perna}\ \emph {et~al.}(2003)\citenamefont {Perna},
  \citenamefont {Narayan}, \citenamefont {Rybicki}, \citenamefont {Stella},\
  and\ \citenamefont {Treves}}]{Perna:2003ck}%
  \BibitemOpen
  \bibfield  {author} {\bibinfo {author} {\bibfnamefont {R.}~\bibnamefont
  {Perna}}, \bibinfo {author} {\bibfnamefont {R.}~\bibnamefont {Narayan}},
  \bibinfo {author} {\bibfnamefont {G.}~\bibnamefont {Rybicki}}, \bibinfo
  {author} {\bibfnamefont {L.}~\bibnamefont {Stella}}, \ and\ \bibinfo {author}
  {\bibfnamefont {A.}~\bibnamefont {Treves}},\ }\href {\doibase 10.1086/377091}
  {\bibfield  {journal} {\bibinfo  {journal} {Astrophys. J.}\ }\textbf
  {\bibinfo {volume} {594}},\ \bibinfo {pages} {936} (\bibinfo {year}
  {2003})},\ \Eprint {http://arxiv.org/abs/astro-ph/0305421}
  {arXiv:astro-ph/0305421 [astro-ph]} \BibitemShut {NoStop}%
%%CITATION = ASTRO-PH/0305421;%%
\bibitem [{\citenamefont {Narayan}\ and\ \citenamefont
  {Yi}(1994)}]{Narayan:1994xi}%
  \BibitemOpen
  \bibfield  {author} {\bibinfo {author} {\bibfnamefont {R.}~\bibnamefont
  {Narayan}}\ and\ \bibinfo {author} {\bibfnamefont {I.-s.}\ \bibnamefont
  {Yi}},\ }\href {\doibase 10.1086/187381} {\bibfield  {journal} {\bibinfo
  {journal} {Astrophys. J.}\ }\textbf {\bibinfo {volume} {428}},\ \bibinfo
  {pages} {L13} (\bibinfo {year} {1994})},\ \Eprint
  {http://arxiv.org/abs/astro-ph/9403052} {arXiv:astro-ph/9403052 [astro-ph]}
  \BibitemShut {NoStop}%
%%CITATION = ASTRO-PH/9403052;%%
\bibitem [{\citenamefont {Narayan}\ and\ \citenamefont
  {McClintock}(2008)}]{Narayan:2008bv}%
  \BibitemOpen
  \bibfield  {author} {\bibinfo {author} {\bibfnamefont {R.}~\bibnamefont
  {Narayan}}\ and\ \bibinfo {author} {\bibfnamefont {J.~E.}\ \bibnamefont
  {McClintock}},\ }\href {\doibase 10.1016/j.newar.2008.03.002} {\bibfield
  {journal} {\bibinfo  {journal} {New Astron. Rev.}\ }\textbf {\bibinfo
  {volume} {51}},\ \bibinfo {pages} {733} (\bibinfo {year} {2008})},\ \Eprint
  {http://arxiv.org/abs/0803.0322} {arXiv:0803.0322 [astro-ph]} \BibitemShut
  {NoStop}%
%%CITATION = ARXIV:0803.0322;%%
\bibitem [{\citenamefont {Tseliakhovich}\ and\ \citenamefont
  {Hirata}(2010)}]{Tseliakhovich:2010bj}%
  \BibitemOpen
  \bibfield  {author} {\bibinfo {author} {\bibfnamefont {D.}~\bibnamefont
  {Tseliakhovich}}\ and\ \bibinfo {author} {\bibfnamefont {C.}~\bibnamefont
  {Hirata}},\ }\href {\doibase 10.1103/PhysRevD.82.083520} {\bibfield
  {journal} {\bibinfo  {journal} {Phys. Rev.}\ }\textbf {\bibinfo {volume}
  {D82}},\ \bibinfo {pages} {083520} (\bibinfo {year} {2010})},\ \Eprint
  {http://arxiv.org/abs/1005.2416} {arXiv:1005.2416 [astro-ph.CO]} \BibitemShut
  {NoStop}%
%%CITATION = ARXIV:1005.2416;%%
\bibitem [{\citenamefont {Hutsi}\ \emph {et~al.}(2009)\citenamefont {Hutsi},
  \citenamefont {Hektor},\ and\ \citenamefont {Raidal}}]{Huetsi:2009ex}%
  \BibitemOpen
  \bibfield  {author} {\bibinfo {author} {\bibfnamefont {G.}~\bibnamefont
  {Hutsi}}, \bibinfo {author} {\bibfnamefont {A.}~\bibnamefont {Hektor}}, \
  and\ \bibinfo {author} {\bibfnamefont {M.}~\bibnamefont {Raidal}},\ }\href
  {\doibase 10.1051/0004-6361/200912760} {\bibfield  {journal} {\bibinfo
  {journal} {Astron. Astrophys.}\ }\textbf {\bibinfo {volume} {505}},\ \bibinfo
  {pages} {999} (\bibinfo {year} {2009})},\ \Eprint
  {http://arxiv.org/abs/0906.4550} {arXiv:0906.4550 [astro-ph.CO]} \BibitemShut
  {NoStop}%
%%CITATION = ARXIV:0906.4550;%%
\bibitem [{\citenamefont {Hutsi}\ \emph {et~al.}(2011)\citenamefont {Hutsi},
  \citenamefont {Chluba}, \citenamefont {Hektor},\ and\ \citenamefont
  {Raidal}}]{Hutsi:2011vx}%
  \BibitemOpen
  \bibfield  {author} {\bibinfo {author} {\bibfnamefont {G.}~\bibnamefont
  {Hutsi}}, \bibinfo {author} {\bibfnamefont {J.}~\bibnamefont {Chluba}},
  \bibinfo {author} {\bibfnamefont {A.}~\bibnamefont {Hektor}}, \ and\ \bibinfo
  {author} {\bibfnamefont {M.}~\bibnamefont {Raidal}},\ }\href {\doibase
  10.1051/0004-6361/201116914} {\bibfield  {journal} {\bibinfo  {journal}
  {Astron. Astrophys.}\ }\textbf {\bibinfo {volume} {535}},\ \bibinfo {pages}
  {A26} (\bibinfo {year} {2011})},\ \Eprint {http://arxiv.org/abs/1103.2766}
  {arXiv:1103.2766 [astro-ph.CO]} \BibitemShut {NoStop}%
%%CITATION = ARXIV:1103.2766;%%
\bibitem [{\citenamefont {Seager}\ \emph {et~al.}(1999)\citenamefont {Seager},
  \citenamefont {Sasselov},\ and\ \citenamefont {Scott}}]{Seager:1999bc}%
  \BibitemOpen
  \bibfield  {author} {\bibinfo {author} {\bibfnamefont {S.}~\bibnamefont
  {Seager}}, \bibinfo {author} {\bibfnamefont {D.~D.}\ \bibnamefont
  {Sasselov}}, \ and\ \bibinfo {author} {\bibfnamefont {D.}~\bibnamefont
  {Scott}},\ }\href {\doibase 10.1086/312250} {\bibfield  {journal} {\bibinfo
  {journal} {Astrophys. J.}\ }\textbf {\bibinfo {volume} {523}},\ \bibinfo
  {pages} {L1} (\bibinfo {year} {1999})},\ \Eprint
  {http://arxiv.org/abs/astro-ph/9909275} {arXiv:astro-ph/9909275 [astro-ph]}
  \BibitemShut {NoStop}%
%%CITATION = ASTRO-PH/9909275;%%
\bibitem [{\citenamefont {Padmanabhan}\ and\ \citenamefont
  {Finkbeiner}(2005)}]{Padmanabhan:2005es}%
  \BibitemOpen
  \bibfield  {author} {\bibinfo {author} {\bibfnamefont {N.}~\bibnamefont
  {Padmanabhan}}\ and\ \bibinfo {author} {\bibfnamefont {D.~P.}\ \bibnamefont
  {Finkbeiner}},\ }\href {\doibase 10.1103/PhysRevD.72.023508} {\bibfield
  {journal} {\bibinfo  {journal} {Phys. Rev.}\ }\textbf {\bibinfo {volume}
  {D72}},\ \bibinfo {pages} {023508} (\bibinfo {year} {2005})},\ \Eprint
  {http://arxiv.org/abs/astro-ph/0503486} {arXiv:astro-ph/0503486 [astro-ph]}
  \BibitemShut {NoStop}%
%%CITATION = ASTRO-PH/0503486;%%
\bibitem [{\citenamefont {Hektor}\ \emph {et~al.}(2018)\citenamefont {Hektor},
  \citenamefont {{H\"utsi}}, \citenamefont {Marzola},\ and\ \citenamefont
  {Vaskonen}}]{Hektor:2018lec}%
  \BibitemOpen
  \bibfield  {author} {\bibinfo {author} {\bibfnamefont {A.}~\bibnamefont
  {Hektor}}, \bibinfo {author} {\bibfnamefont {G.}~\bibnamefont {{H\"utsi}}},
  \bibinfo {author} {\bibfnamefont {L.}~\bibnamefont {Marzola}}, \ and\
  \bibinfo {author} {\bibfnamefont {V.}~\bibnamefont {Vaskonen}},\ }\href@noop
  {} {\  (\bibinfo {year} {2018})},\ \Eprint {http://arxiv.org/abs/1805.09319}
  {arXiv:1805.09319 [hep-ph]} \BibitemShut {NoStop}%
%%CITATION = ARXIV:1805.09319;%%
\bibitem [{\citenamefont {Tisserand}\ \emph {et~al.}(2007)\citenamefont
  {Tisserand} \emph {et~al.}}]{Tisserand:2006zx}%
  \BibitemOpen
  \bibfield  {author} {\bibinfo {author} {\bibfnamefont {P.}~\bibnamefont
  {Tisserand}} \emph {et~al.} (\bibinfo {collaboration} {EROS-2}),\ }\href
  {\doibase 10.1051/0004-6361:20066017} {\bibfield  {journal} {\bibinfo
  {journal} {Astron. Astrophys.}\ }\textbf {\bibinfo {volume} {469}},\ \bibinfo
  {pages} {387} (\bibinfo {year} {2007})},\ \Eprint
  {http://arxiv.org/abs/astro-ph/0607207} {arXiv:astro-ph/0607207 [astro-ph]}
  \BibitemShut {NoStop}%
%%CITATION = ASTRO-PH/0607207;%%
\bibitem [{\citenamefont {Allsman}\ \emph {et~al.}(2001)\citenamefont {Allsman}
  \emph {et~al.}}]{Allsman:2000kg}%
  \BibitemOpen
  \bibfield  {author} {\bibinfo {author} {\bibfnamefont {R.~A.}\ \bibnamefont
  {Allsman}} \emph {et~al.} (\bibinfo {collaboration} {Macho}),\ }\href
  {\doibase 10.1086/319636} {\bibfield  {journal} {\bibinfo  {journal}
  {Astrophys. J.}\ }\textbf {\bibinfo {volume} {550}},\ \bibinfo {pages} {L169}
  (\bibinfo {year} {2001})},\ \Eprint {http://arxiv.org/abs/astro-ph/0011506}
  {arXiv:astro-ph/0011506 [astro-ph]} \BibitemShut {NoStop}%
%%CITATION = ASTRO-PH/0011506;%%
\bibitem [{\citenamefont {Zumalacarregui}\ and\ \citenamefont
  {Seljak}(2017)}]{Zumalacarregui:2017qqd}%
  \BibitemOpen
  \bibfield  {author} {\bibinfo {author} {\bibfnamefont {M.}~\bibnamefont
  {Zumalacarregui}}\ and\ \bibinfo {author} {\bibfnamefont {U.}~\bibnamefont
  {Seljak}},\ }\href@noop {} {\  (\bibinfo {year} {2017})},\ \Eprint
  {http://arxiv.org/abs/1712.02240} {arXiv:1712.02240 [astro-ph.CO]}
  \BibitemShut {NoStop}%
%%CITATION = ARXIV:1712.02240;%%
\bibitem [{\citenamefont {Koushiappas}\ and\ \citenamefont
  {Loeb}(2017)}]{Koushiappas:2017chw}%
  \BibitemOpen
  \bibfield  {author} {\bibinfo {author} {\bibfnamefont {S.~M.}\ \bibnamefont
  {Koushiappas}}\ and\ \bibinfo {author} {\bibfnamefont {A.}~\bibnamefont
  {Loeb}},\ }\href {\doibase 10.1103/PhysRevLett.119.041102} {\bibfield
  {journal} {\bibinfo  {journal} {Phys. Rev. Lett.}\ }\textbf {\bibinfo
  {volume} {119}},\ \bibinfo {pages} {041102} (\bibinfo {year} {2017})},\
  \Eprint {http://arxiv.org/abs/1704.01668} {arXiv:1704.01668 [astro-ph.GA]}
  \BibitemShut {NoStop}%
%%CITATION = ARXIV:1704.01668;%%
\bibitem [{\citenamefont {Brandt}(2016)}]{Brandt:2016aco}%
  \BibitemOpen
  \bibfield  {author} {\bibinfo {author} {\bibfnamefont {T.~D.}\ \bibnamefont
  {Brandt}},\ }\href {\doibase 10.3847/2041-8205/824/2/L31} {\bibfield
  {journal} {\bibinfo  {journal} {Astrophys. J.}\ }\textbf {\bibinfo {volume}
  {824}},\ \bibinfo {pages} {L31} (\bibinfo {year} {2016})},\ \Eprint
  {http://arxiv.org/abs/1605.03665} {arXiv:1605.03665 [astro-ph.GA]}
  \BibitemShut {NoStop}%
%%CITATION = ARXIV:1605.03665;%%
\bibitem [{\citenamefont {Monroy-Rodr{\'\i}guez}\ and\ \citenamefont
  {Allen}(2014)}]{Monroy-Rodriguez:2014ula}%
  \BibitemOpen
  \bibfield  {author} {\bibinfo {author} {\bibfnamefont {M.~A.}\ \bibnamefont
  {Monroy-Rodr{\'\i}guez}}\ and\ \bibinfo {author} {\bibfnamefont
  {C.}~\bibnamefont {Allen}},\ }\href {\doibase 10.1088/0004-637X/790/2/159}
  {\bibfield  {journal} {\bibinfo  {journal} {Astrophys. J.}\ }\textbf
  {\bibinfo {volume} {790}},\ \bibinfo {pages} {159} (\bibinfo {year}
  {2014})},\ \Eprint {http://arxiv.org/abs/1406.5169} {arXiv:1406.5169
  [astro-ph.GA]} \BibitemShut {NoStop}%
%%CITATION = ARXIV:1406.5169;%%
\bibitem [{\citenamefont {Raidal}\ \emph {et~al.}(2017)\citenamefont {Raidal},
  \citenamefont {Vaskonen},\ and\ \citenamefont
  {Veerm{\"a}e}}]{Raidal:2017mfl}%
  \BibitemOpen
  \bibfield  {author} {\bibinfo {author} {\bibfnamefont {M.}~\bibnamefont
  {Raidal}}, \bibinfo {author} {\bibfnamefont {V.}~\bibnamefont {Vaskonen}}, \
  and\ \bibinfo {author} {\bibfnamefont {H.}~\bibnamefont {Veerm{\"a}e}},\
  }\href {\doibase 10.1088/1475-7516/2017/09/037} {\bibfield  {journal}
  {\bibinfo  {journal} {JCAP}\ }\textbf {\bibinfo {volume} {1709}},\ \bibinfo
  {pages} {037} (\bibinfo {year} {2017})},\ \Eprint
  {http://arxiv.org/abs/1707.01480} {arXiv:1707.01480 [astro-ph.CO]}
  \BibitemShut {NoStop}%
%%CITATION = ARXIV:1707.01480;%%
\bibitem [{\citenamefont {Nakamura}\ \emph {et~al.}(1997)\citenamefont
  {Nakamura}, \citenamefont {Sasaki}, \citenamefont {Tanaka},\ and\
  \citenamefont {Thorne}}]{Nakamura:1997sm}%
  \BibitemOpen
  \bibfield  {author} {\bibinfo {author} {\bibfnamefont {T.}~\bibnamefont
  {Nakamura}}, \bibinfo {author} {\bibfnamefont {M.}~\bibnamefont {Sasaki}},
  \bibinfo {author} {\bibfnamefont {T.}~\bibnamefont {Tanaka}}, \ and\ \bibinfo
  {author} {\bibfnamefont {K.~S.}\ \bibnamefont {Thorne}},\ }\href {\doibase
  10.1086/310886} {\bibfield  {journal} {\bibinfo  {journal} {Astrophys. J.}\
  }\textbf {\bibinfo {volume} {487}},\ \bibinfo {pages} {L139} (\bibinfo {year}
  {1997})},\ \Eprint {http://arxiv.org/abs/astro-ph/9708060}
  {arXiv:astro-ph/9708060 [astro-ph]} \BibitemShut {NoStop}%
%%CITATION = ASTRO-PH/9708060;%%
\bibitem [{\citenamefont {Clark}\ \emph {et~al.}(2018)\citenamefont {Clark},
  \citenamefont {Dutta}, \citenamefont {Gao}, \citenamefont {Ma},\ and\
  \citenamefont {Strigari}}]{Clark:2018ghm}%
  \BibitemOpen
  \bibfield  {author} {\bibinfo {author} {\bibfnamefont {S.}~\bibnamefont
  {Clark}}, \bibinfo {author} {\bibfnamefont {B.}~\bibnamefont {Dutta}},
  \bibinfo {author} {\bibfnamefont {Y.}~\bibnamefont {Gao}}, \bibinfo {author}
  {\bibfnamefont {Y.-Z.}\ \bibnamefont {Ma}}, \ and\ \bibinfo {author}
  {\bibfnamefont {L.~E.}\ \bibnamefont {Strigari}},\ }\href@noop {} {\
  (\bibinfo {year} {2018})},\ \Eprint {http://arxiv.org/abs/1803.09390}
  {arXiv:1803.09390 [astro-ph.HE]} \BibitemShut {NoStop}%
%%CITATION = ARXIV:1803.09390;%%
\bibitem [{\citenamefont {Slatyer}\ and\ \citenamefont
  {Wu}(2017)}]{Slatyer:2016qyl}%
  \BibitemOpen
  \bibfield  {author} {\bibinfo {author} {\bibfnamefont {T.~R.}\ \bibnamefont
  {Slatyer}}\ and\ \bibinfo {author} {\bibfnamefont {C.-L.}\ \bibnamefont
  {Wu}},\ }\href {\doibase 10.1103/PhysRevD.95.023010} {\bibfield  {journal}
  {\bibinfo  {journal} {Phys. Rev.}\ }\textbf {\bibinfo {volume} {D95}},\
  \bibinfo {pages} {023010} (\bibinfo {year} {2017})},\ \Eprint
  {http://arxiv.org/abs/1610.06933} {arXiv:1610.06933 [astro-ph.CO]}
  \BibitemShut {NoStop}%
%%CITATION = ARXIV:1610.06933;%%
\end{thebibliography}%

\end{document}